\documentclass[aps,prb,twocolumn,letterpaper]{revtex4}
\usepackage{graphicx}
\usepackage{dcolumn}
\usepackage{amsmath}
\usepackage{amssymb}
\usepackage{subfigure}
\usepackage{float}
\usepackage{epstopdf}
\usepackage{natbib}
\def\kv{{\bf k}}
\def\qv{{\bf q}}

\begin{document}

\title{Effects of order parameter self-consistency in a s$\pm$-$s$ junction}
\author{Rosa Rodr\'{i}guez-Mota$^1$, Erez Berg$^2$ and T. Pereg-Barnea$^1$}
\affiliation{$^1$Department of Physics and the Centre for Physics of Materials, McGill University, Montreal, Quebec,
Canada H3A 2T8\\
$^2$Department of Condensed Matter Physics, Weizmann Institute of Science, Rehovot, Israel 76100}
\date{\today}
\newcommand{\etalcomma}{\textit{et al.,\ }}
\newcommand{\average}[2]{ \left\langle #1 #2 \right\rangle }
\newcommand{\creation}[2]{ #1_{#2}^{\dagger}}
\newcommand{\destruction}[2]{ #1_{#2}}
\newcommand{\MatEl}[1]{ \left(#1\right)_{m,n}}

\begin{abstract}
The properties of Josephson tunneling between a single band $s$-wave superconductor and a two band $s_\pm$ superconductor are studied, in relation to recent experiments involving iron-based superconductors. We study both a single junction and a loop consisting of two junctions. In both cases, the relative phase between the order parameters of the two superconductors is tuned and the energy of the system is calculated. In a single junction, we find four types of behaviors characterized by the location of minima in the energy/phase relations. These phases include a newly found double minimum junction which appears only when the order parameters are treated self-consistently. We analyze the loop geometry setup in light of our results for a single junction, where the phase difference in the junctions is controlled by a threaded flux. We find four types of energy/flux relations. These include states for which the energy is minimized when the threaded flux is an integer or half integer number of flux quanta, a time reversal broken state and a meta-stable state.
\end{abstract}

\maketitle

\section{Introduction}
The experimental determination of the pairing symmetry of an unconventional superconductor is an important tool for narrowing down the microscopic theories which suggest different origins for superconductivity in the system. Shortly after the discovery of the iron-based superconductors (FeSCs)\citep{FeSC2006,FeSC2008}, spin fluctuations were proposed as the pairing mechanism in this family. This mechanism results in a novel pairing symmetry, called $s_\pm$\citep{symmetryrev1,symmetryrev2}. The $s_\pm$ order parameter is finite on both the hole and electron Fermi pockets but changes sign between the Brillouin zone $\Gamma$ point, where the hole pockets are, and the $M$ point, where the electron pockets reside. Although most evidence point towards spin fluctuations as the pairing mechanism, superconductivity on FeSCs could also arise from orbital fluctuations, in which case, the order parameter on electron and hole pockets would have the same phase\citep{symmetryrev1,symmetryrev2}. It is therefore crucial to pin down the possible sign difference between the two types of pockets. The sign difference has proven challenging to detect\citep{Moler,Hoffman,VortexHoffman1,VortexHoffman2,Josephsoncaxis,FeJosephsonreview,
pointcontactspectroscopyreview,PhysRevB.91.214501}, and despite experimental evidence in favor of $s_\pm$\citep{neutronscat1,neutronscat2,QPI,chen2010}, the pairing symmetry of FeSCs has not been unequivocally determined.

One important tool for detecting the order parameter structure is the Josephson effect, due to its sensitivity to the order parameter (OP) phase difference across the junction.
The Josephson effect played a key role in determining the $d$-wave nature of the order parameter of the high T$_{\rm c}$ cuprate superconductors\citep{cupratereview,PhysRevLett.71.2134}. There, the phase of the order parameter is tied to the crystallographic direction and one can engineer a $\pi$ corner junction by piecing together samples in different orientations. In contrast, identifying a sign change in the case of the iron based superconductors is more challenging. This is because the sign change is expected between Fermi pockets at low momenta (at the $\Gamma$ point) and Fermi pockets at high momenta (at the $M$ point). Therefore, a rotation of the physical lattice does not result in a sign change.

In iron based superconductors (FeSCs), some evidence in favor of $s_\pm$ OP was provided by a loop-flux experiment by Chen {\it et al.}\citep{chen2010}. The setup consisted of a niobium fork making two contacts with a sample of $\mathrm{NdFeAsO_{0.88}F_{0.12}}$. This amounts to a loop made of a conventional $s$-wave superconductor which is connected in two points to an FeSC sample.
This loop was subjected to a pulse of magnetic flux after which the flux in the loop was measured over time. Flux jumps of integer and half integer units of the superconducting flux quantum were observed. As explained in Ref.~[\onlinecite{metastablepijunction}], this can be interpreted as a meta-stable $1/2$-flux loop, possible in the case of $s_\pm$-$s$-wave loop.

The problem of a junction between an $s$-wave superconductor and an $s_\pm$-superconductor was considered by several authors previously\citep{quasiclassical,quasiclassical2,Microscopic2,multibandjunctions-sdw,secondharmonicballistic,metastablepijunction,TRBGL1,TRB2,Lagrangian,secondharmonicballisticprevious,
Microscopic2previous,wire,temperature,trilayerproposal,PhysRevB.66.214507,GLold,PhysRevB.91.214501}. Those include different approaches such as the Ginzburg-Landau (GL) formalism\citep{TRBGL1,metastablepijunction,GLold}, calculating Josephson current from Andreev levels\citep{secondharmonicballistic}, or through Usadel quasiclassical equation\citep{secondharmonicballisticprevious,wire,Microscopic2previous}, among others. 

The literature points at two possible types of contacts: (i) the $s$-wave superconductor couples predominantly to either the electron or hole pockets or (ii) the couplings between the $s$-wave and the electron and hole pockets are comparable leading to Josephson frustration. One type of proposals to experimentally determine the $s_\pm$ symmetry rely on the ability to produce different types of contacts in which the $s$-wave predominantly couples to one type of pocket or another\citep{trijunctionpointcontact,secondharmonicballistic,secondharmonicballisticprevious,GolubovandMazin,ExperimentProposal_BarierThickness,cornerjunctionmodification}. Other experimental proposals assume comparable coupling to both pockets and conclude that the Josephson frustration can lead to a time reversal symmetry breaking phase (TRB) in a loop setup\citep{metastablepijunction,Microscopic2,Lagrangian,TRBGL1,TRB2}.

It has been argued\citep{Microscopic2,secondharmonicballistic,PhysRevB.91.214501}, and will be argued in this work, that higher order harmonics in the Josephson current, which are often neglected, for instance within the Ginzburg-Landau\citep{TRBGL1,metastablepijunction,GLold} and the Usadel quasiclassical equation\citep{Usadel,secondharmonicballisticprevious,wire,Microscopic2previous} approaches, become very important in this scenario. At low temperatures, this is especially crucial in the case of comparable coupling between the $s$-wave superconductor and the hole/electron pockets.

In the current paper we work with a microscopic model on a lattice in an $s_\pm$-$s$-wave junction setup. While our model is similar to that investigated by previous authors\cite{quasiclassical,quasiclassical2,PhysRevB.91.214501} our treatment is different as we solve the Bogoliubov deGennes equations {\it self-consistently}. The self-consistently causes the order parameters of the superconductors on both sides of the junction to be a function of the distance from the junction, both their amplitude and phase. This helps the system relieve some of its Josephson frustration and leads to important differences from the non self-consistent treatment. Namely, we find that a double minimum structure in the energy/phase difference relation is obtained only when the Bogoliubov-deGennes equations are solved self-consistently.

In the next section, we present our model and method for a single $s_\pm$-$s$-wave junction (section~\ref{sec:JJmodel}). The results are presented in section~\ref{sec:JJresults} and discussed in section~\ref{sec:JJorigin}. Section~\ref{sec:loop} discusses the combination of two $s_\pm$-$s$ junctions into a loop and its possible states.

\section{$s_\pm$-$s$ junction}
\subsection{The model}\label{sec:JJmodel}

\begin{figure}
\centering
\includegraphics[width=\linewidth]{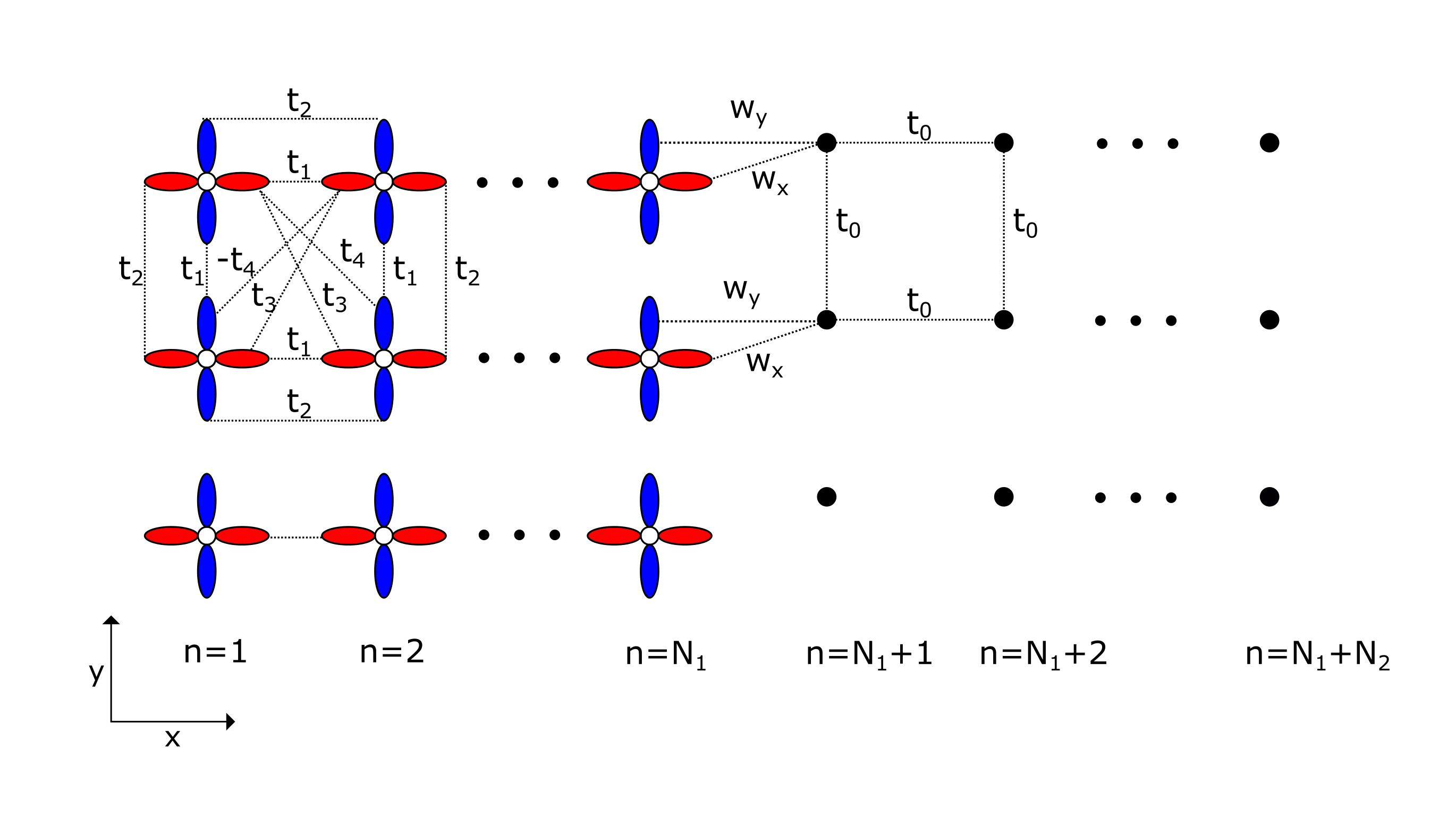}
\caption{Illustration of the planar junction studied in this work showing the different hopping parameters. The $s_\pm$ superconductor (left) has two orbitals per site and the $s$ superconductor (right) one.}
\label{fig:junction}
\end{figure}

We study the Josephson junction depicted in Fig.~\ref{fig:junction} within a tight-binding formalism. In this arrangement, we consider both superconductors to be two dimensional, and the tunneling between them is planar and directed along the (1 0) direction. For simplicity, the lattice constant of both superconductors is taken to be equal and will be set to 1 for the remainder of this paper. In order to provide a better understanding of our model, we first write the Hamiltonian for each superconductor separately without tunneling between them.

\subsubsection*{$s_\pm$ Superconductor}
For the $s_\pm$ superconductor, we use a minimal two orbital model\citep{2bandmodel} in which the two orbitals correspond to the $3d_{xz}$ and $3d_{yz}$ iron orbitals illustrated by red/blue lobes in Fig.~\ref{fig:junction}. In this model, there are four different types of hopping: $t_1$ is the amplitude of nearest neighbor intra-orbital hopping in the direction in which the orbitals maximally overlap, $t_2$ is the nearest neighbor intra-orbital hopping amplitude in the direction in which the orbitals minimally overlap, $t_3$ is the next-nearest neighbor intra-orbital hopping amplitude, and $t_4$ is the next-nearest neighbor inter-orbital hopping. We add Cooper pairing with $s_\pm$ symmetry in the form of an intra-orbital pairing\citep{J1J2model,quasiclassical2} with the momentum structure $\cos k_x \cos k_y$. We define the operators $\creation{d}{x,\mathbf{k},\sigma}{}$ and $\creation{d}{y,\mathbf{k},\sigma}{}$ which create an electron in the $d_{xz}$, $d_{yz}$ orbital with momentum $\mathbf{k}$ and spin $\sigma$. Using these operators we write the Hamiltonian $H_{s_\pm}$ in the form $H_{s_\pm}= \sum_\mathbf{k} \Psi^\dagger \left( \mathbf{k} \right) A \left( \mathbf{k} \right) \Psi \left( \mathbf{k} \right)$, where $\Psi^{\dagger} \left(\mathbf{k}\right) = \left( \creation{d}{x,\mathbf{k},\uparrow}{},\destruction{d}{x,-\mathbf{k},\downarrow}{} ,\creation{d}{y,\mathbf{k},\uparrow}{},\destruction{d}{y,-\mathbf{k},\downarrow}{} \right)$, and
\begin{equation}
A \left( \mathbf{k} \right) = \left( \begin{array}{cccc}
\epsilon_x \left( \mathbf{k} \right) & \Delta_x \left(\mathbf{k}\right) & \epsilon_{xy} \left(\mathbf{k} \right) & 0 \\
\Delta_x^\ast \left(\mathbf{k}\right) & -\epsilon_x \left( \mathbf{k} \right) & 0 & -\epsilon_{xy} \left(\mathbf{k} \right)\\
\epsilon_{xy} \left(\mathbf{k} \right) & 0 & \epsilon_y \left( \mathbf{k} \right) & \Delta_y \left(\mathbf{k}\right) \\
0 & -\epsilon_{xy} \left(\mathbf{k} \right) & \Delta_y^\ast \left(\mathbf{k}\right) & -\epsilon_y \left( \mathbf{k} \right) \\
\end{array} \right)
\label{eqn:spmMF}
\end{equation}
\begin{align}
\begin{split}
\epsilon_{x} \left(\mathbf{k}\right) &= -2 t_1 \cos k_x -2 t_2 \cos k_y - 4 t_3 \cos k_x \cos k_y -\mu \\
\epsilon_{y} \left(\mathbf{k}\right) &= -2 t_2 \cos k_x -2 t_1 \cos k_y - 4 t_3 \cos k_x \cos k_y - \mu\\
\epsilon_{xy} \left(\mathbf{k}\right) &= -4 t_4 \sin k_x \sin k_y \\
\Delta_{x,y} \left( \mathbf{k} \right) &= \Delta_{x,y} \cos k_x \cos k_y
\label{eqn:spmParams}
\end{split}
\end{align}
where $\mu$ is the chemical potential, $\Delta_{x,y}$ is the pairing amplitude and $\epsilon_i$ are the Fourier transforms of the hoping amplitudes.

Close to the junction we expect the order parameter of both superconductors to become dependent on position. A mean-field Hamiltonian cannot account for the effects of this order parameter modification, therefore it is necessary to write down interaction terms which lead to superconductivity in our system. The position dependence of the order parameter is then determined self-consistently.

We consider the $s_\pm$ order parameter to arise from a $J_1-J_2$ nearest neighbors and next nearest neighbors anti-ferrromagnetic Heisenberg interaction. The mean-field decoupling of this interaction leads to four possible pairing symmetries. The dominant pairing symmetry for this model is $s_\pm$ in the relevant region of parameters\citep{J1J2model}. For simplicity, we only consider the terms of the interaction that lead to an intra-orbital $s_\pm$ pairing. Hence we write the interaction term as:
\begin{equation}\label{eq:interaction1}
\sum_{\kv,\kv',\qv,\alpha} V\left(\kv,\kv',\qv\right)\creation{d}{\alpha,\kv,\uparrow}{} \creation{d}{\alpha,-\kv+\qv,\downarrow}{} \destruction{d}{\alpha,-\kv'+\qv,\downarrow}{} \destruction{d}{\alpha,\kv',\uparrow}{}
\end{equation}
where
\begin{equation}\label{eq:interaction2}
\begin{split}
V\left(\kv,\kv',\qv\right) = -\frac{8 J_2}{N} \cos \left( k_x -\frac{q_x}{2} \right) \cos \left( k_y -\frac{q_y}{2} \right) \times \\
 \cos \left( k'_x -\frac{q_x}{2} \right) \cos \left( k'_y -\frac{q_y}{2} \right),
\end{split}
\end{equation}
and $\alpha = x,y$ here and throughout the paper. This interaction is decoupled as:
\begin{equation}
 \sum_{\kv,\qv,\alpha} \Delta_{\alpha}(\qv) f\left(\kv,\qv\right) \creation{d}{\alpha,\kv,\uparrow}{} \creation{d}{\alpha,-\kv+\qv,\downarrow}{} + h.c.
\end{equation}
with the structure factor
\begin{equation}
f\left(\kv,\qv \right) = \cos \left( k_x -\frac{q_x}{2} \right) \cos \left( k_y -\frac{q_y}{2} \right),
\end{equation}
and
\begin{equation}
\Delta_{\alpha}(\qv) = -\frac{8 J_2}{N} \sum_{\kv'} f\left(\kv',\qv \right) \left\langle \destruction{d}{\alpha,-\kv'+\qv,\downarrow}{} \destruction{d}{\alpha,\kv',\uparrow}{} \right\rangle.
\label{eqn:SCspm}
\end{equation}
In a system with translation invariance, the ground state corresponds to zero momentum pairing, i.e. $\Delta_{\alpha}(\qv) = \Delta_{\alpha} \delta_{\qv,0}$ resulting in the mean field Hamiltonian, Eq.~(\ref{eqn:spmMF}). Furthermore, the self-consistent solution in the translational invariant system gives $\Delta_{x} = \Delta_{y}$, which corresponds to electron-electron and hole-hole pairing with opposite signs. 

\subsubsection*{$s$-wave Superconductor}

The $s$-wave superconductor is modeled using one orbital per site, nearest neighbors hopping $t_0$, and momentum independent pairing $\Delta_0$. The operator $c^\dagger_{\sigma \kv}$ creates an electron with momentum $\kv$ and spin $\sigma$ on the $s$-wave superconductor side. The system is described by $H_s = \sum_{\kv} \Phi_\kv^\dagger B_\kv \Phi_\kv$, where $\Phi_\kv^\dagger=\left( \begin{array}{cc}
\creation{c}{\kv,\uparrow}{} & \destruction{c}{-\kv,\downarrow}{} \end{array} \right)$,
\begin{equation}
B_\kv = \left(
\begin{array}{c c} \epsilon_0 \left(\kv\right) - \mu_0 & \Delta_0 \\
\Delta_0^* & -\epsilon_0 \left(\kv\right) + \mu_0 \end{array} \right),
\end{equation}
$\epsilon_0 \left(\kv\right) = -2 t_0\left( \cos \left( k_x \right) + \cos \left(k_y\right)\right)$, $\mu_0$ is the chemical potential, and $\Delta_0$ the pairing amplitude.

To account for superconductivity in the $s$-wave side of the system, we use an attractive Hubbard-$U$ term:
\begin{equation}\label{eq:interaction3}
-\frac{U}{N}\sum_{\kv,\kv',\alpha} \creation{c}{\kv,\uparrow}{} \creation{c}{-\kv+\qv,\downarrow}{} \destruction{c}{-\kv'+\qv,\downarrow}{} \destruction{c}{\kv',\uparrow}{},
\end{equation}
with $U>0$. In a translation-invariant system, the mean-field decoupling of this interaction leads to the Hamiltonian $H_s$ mentioned above and the self-consistency equation:
\begin{equation}
\Delta_0 = -\frac{U}{N} \sum_{\kv} \left\langle \destruction{c}{-\kv,\downarrow}{} \destruction{c}{\kv,\uparrow}{} \right\rangle.
\label{eqn:SCs}
\end{equation}

\subsubsection*{Junction}

The model of the $s_\pm$-$s$ junction considered in this work consists of the two superconductors previously described connected through a tunneling contact along the (1 0) direction. We consider $N_1$ lattice sites in the $x$-direction in the $s_\pm$ superconductor and $N_2$ in the $s$-wave superconductor. The contact breaks the translation symmetry along the $x$-direction, but since the $y$-direction is still periodic, the momentum $k_y$ is well defined. The natural description of the system is in terms of the operator $\creation{d}{\alpha,k_y,\sigma}(n)$ which create an electron whose momentum component in the $y$-direction is $k_y$, on a site with $x$-coordinate index $n$, with spin $\sigma$ in the $\alpha$ orbital, and $\creation{c}{k_y,\sigma}(n)$ which creates an electron whose momentum component in the $y$-direction is $k_y$, on a site with $x$-coordinate $n$, with spin $\sigma$ in the $s$-wave superconductor. In order to write the Hamiltonian, we define the vectors
\begin{equation}
\begin{split}
d_{\alpha,\sigma}^{\dagger} (k_y) &= \left(\begin{array}{ccc}
d_{\alpha,k_y,\sigma}^{\dagger} (1) & ... & d_{\alpha,k_y,\sigma}^{\dagger} (N_1)
\end{array}\right),\mathrm{and}\\
c_{\sigma}^{\dagger} (k_y) &= \left(\begin{array}{ccc}
c_{k_y,\sigma}^{\dagger} (N_1+1) & ... & c_{k_y,\sigma}^{\dagger} (N_1+N_2)
\end{array}\right)
\end{split}
\end{equation}
which contain all the possible creation operators for a given $k_y$, spin and orbital. Combining them into Nambu vectors
\begin{equation}
\begin{split}
&\Psi_{s_\pm}^\dagger (k_y) = \left(\begin{array}{cccc}
d_{x,\uparrow}^{\dagger} (k_y) & d_{y,\uparrow}^{\dagger} (k_y) & d_{x,\downarrow}^{T} (-k_y)& d_{y,\downarrow}^{T} (-k_y)
\end{array}\right) \\
&\Psi_s^\dagger(k_y) = \left(\begin{array}{cc}
c_{\uparrow}^{\dagger} (k_y) & c_{\downarrow}^T (-k_y)
\end{array}\right),
\end{split}
\end{equation}
allows us to write the Hamiltonian in the following compact form:
\begin{equation}
H = \sum_{k_y} \Psi^{\dagger}(k_y)
\left( \begin{array}{cc}
H_{s_{\pm}}(k_y) & T\\
T^\dagger & H_s(k_y)
\end{array} \right) \Psi(k_y) + C,
\label{eqn:juncHamiltonian}
\end{equation}
where $C$ is defined below and 
\begin{equation*}
\Psi^\dagger(k_y)= \left(\begin{array}{cccc}
\Psi_{s_\pm}^\dagger (k_y) & \Psi_s^\dagger(k_y)
\end{array}\right).
\end{equation*}
The matrices $H_{s_{\pm}}(k_y)$ and $H_{s}(k_y)$ are given by the following BCS form:
\begin{equation}
H_{s_{\pm}(s)}(k_y) = \left( \begin{array}{cc}
K_{s_{\pm}(s)}(k_y) & \Delta_{s_{\pm}(s)}(k_y) \\
\Delta_{s_{\pm}(s)}(k_y)^\dagger & -K_{s_{\pm}(s)}(-k_y)^*
\end{array} \right)
\end{equation}

For the $s$-wave part of the Hamiltonian $H_{s}(k_y)$, $K_s (k_y)$ and $\Delta_{s} (k_y)$, are $N_2 \times N_2$ matrices given by:
\begin{align}
\begin{split}
\MatEl{\Delta_{s} (k_y)}= &\Delta_{s} (N_1+n) \delta_{m,n} \\
\MatEl{K_s (k_y)} =& -\left(2t_0 \cos(k_y) + \mu_0\right) \delta_{m,n}\\& -t_0 \left( \delta_{m,n+1} + \delta_{m,n-1} \right).
\end{split}
\end{align}

For the $s_\pm$ part of the Hamiltonian the matrices $K_{s_{\pm}}(k_y)$ and $\Delta_{s_{\pm}(s)}(k_y)$ can be further decomposed as:
\begin{equation}
K_{s_{\pm}}(k_y) = \left( \begin{array}{cc}
K_x(k_y) & K_{xy}(k_y) \\
K_{xy}(k_y) & K_y(k_y)
\end{array} \right),
\end{equation}
and
\begin{equation}
\Delta_{s_{\pm}}(k_y) = \left( \begin{array}{cc}
\Delta_x(k_y) & 0 \\
0 & \Delta_y(k_y)
\end{array} \right),
\end{equation}
where the above sub-blocks are the following $N_1 \times N_1$ matrices:
\begin{align}
\begin{split}
\MatEl{K_x(k_y)} =& - \left( 2t_2 \cos(k_y) + \mu \right) \delta_{m,n} \\ & - \left(t_1 + 2 t_3 \cos(k_y)\right)\left( \delta_{m,n+1} + \delta_{m,n-1} \right) \\
\MatEl{K_y(k_y)} =& - \left( 2t_1 \cos(k_y) + \mu \right) \delta_{m,n} \\ & - \left(t_2 + 2 t_3 \cos(k_y)\right)\left( \delta_{m,n+1} + \delta_{m,n-1} \right) \\
\MatEl{K_{xy}(k_y)} =& -2 i t_4 \sin (k_y) \left( \delta_{m,n+1} - \delta_{m,n-1} \right) \\
\MatEl{\Delta_{\alpha}(k_y)} = & \Delta_{\alpha}(n+1,n) \cos(k_y) \delta_{m,n+1} \\& +\Delta_{\alpha}(n-1,n) \cos(k_y) \delta_{m,n-1}
\end{split}
\end{align}

The matrix $T$ describes the tunneling contact between the two superconductors. As shown in Figure \ref{fig:junction}, we consider hopping an electron in the $3d_{xz}$ ($3d_{yz}$) orbital of the last site of the $s_\pm$ superconductor to the first site of the $s$-wave superconductor with an amplitude $w_x$ ($w_y$). Hence, $T$ is given by:

\begin{equation}
T = \left(\begin{array}{cc}
T_x & 0 \\
T_y & 0 \\
0 & -T_x \\
0 & -T_y \\
\end{array} \right),
\end{equation}
with,
\begin{align}
\begin{split}
\MatEl{T_{\alpha}} = -w_{\alpha}\delta_{m,N_1} \delta_{n,1}\\
\end{split}
\end{align}
here $T_x$ and $T_y$ are $N_1 \times N_2$ matrices.

Finally $C$ in Eq.~(\ref{eqn:juncHamiltonian}) is equal to:
\small
\begin{equation}
C = \frac{N_y}{2 J_2} \sum_{\alpha,n=1}^{N_1-1}|\Delta_\alpha (n,n+1)|^2 -\frac{N_y}{U} \sum_{n=N_1+1}^{N_1+N_2} |\Delta_0 (n)|^2
\end{equation}
\normalsize
Since superconductivity arises from the spin interaction terms given in Eqs.~\ref{eq:interaction1}-\ref{eq:interaction2},\ref{eq:interaction3}, the following self-consistency equations should be satisfied:
\begin{equation}
\begin{split}
\Delta_{\alpha} (n,n+1)= & -\frac{2 J_2}{N_y} \sum_{k_y} \cos k_y \left(
g_{\alpha,k_y}\left(n+1,n\right) \right. \\ &\left. + g_{\alpha,k_y}\left(n,n+1\right) \right) \\
\Delta_{s} (n) &= -\frac{U}{N_y} \sum_{k_y} \left\langle c_{-k_y,\downarrow} (n) c_{k_y,\uparrow} (n) \right\rangle ,
\end{split}
\label{eqn:sc}
\end{equation}
with $g_{\alpha,k_y}\left(m,n\right)= \left\langle d_{\alpha,-k_y,\downarrow} (m) d_{\alpha,k_y,\uparrow} (n) \right\rangle$.

We study how the energy of the system and the current depend on the phase difference between the two superconductors.
In an infinite system, this can be modeled by fixing the order parameter at $\pm \infty $ and imposing a phase difference between the two ends. The self-consistency equations of our lattice model are complicated and must be solved numerically, on a finite lattice.
Therefore, we model the composite system by dividing each superconductor into a bulk part and a junction.
In the bulk of the $s_\pm$ superconductor, we set $\Delta_x (n,n+1)= \Delta_y(n,n+1)= \Delta_{\pm}$, where $\Delta_{\pm}$ is real, positive and equal to the pairing amplitude that is obtained self-consistently in a translation invariant system, ignoring the contact. In the bulk part of the $s$-wave superconductor, $\Delta_0(n) = |\Delta_0| e^{i \phi}$, with $|\Delta_0|$ self-consistently determined in the absence of the interface. 
$\phi$ is the phase difference between the $s$-wave order parameter away from the contact and the $s_\pm$ order parameter away from the contact on the other side. In the part near the junction, $\Delta_x (n,n+1)$, $\Delta_y(n,n+1)$ and $\Delta_0(n)$ are determined by the self-consistency equations, Eq.~(\ref{eqn:sc}). Once the order parameters of the system are determined self-consistently, the energy of the system can be found. The current across the contact can be obtained from the energy dependence on the phase difference as $I \left( \phi \right) = \frac{2\mathrm{e}}{\hbar}\frac{dE}{d\phi}$. For simplicity, we ignore the small phase gradient in the bulk in situations where a supercurrent is flowing through the junction. 

\subsection{Results}\label{sec:JJresults}

\begin{figure}
\centering
\begin{minipage}[c]{0.48\linewidth}
\textbf{Self-consistent}
\end{minipage}
\begin{minipage}[c]{0.48\linewidth}
\textbf{Non self-consistent}
\end{minipage}
\subfigure[\ ]{\includegraphics[width=0.48\linewidth]{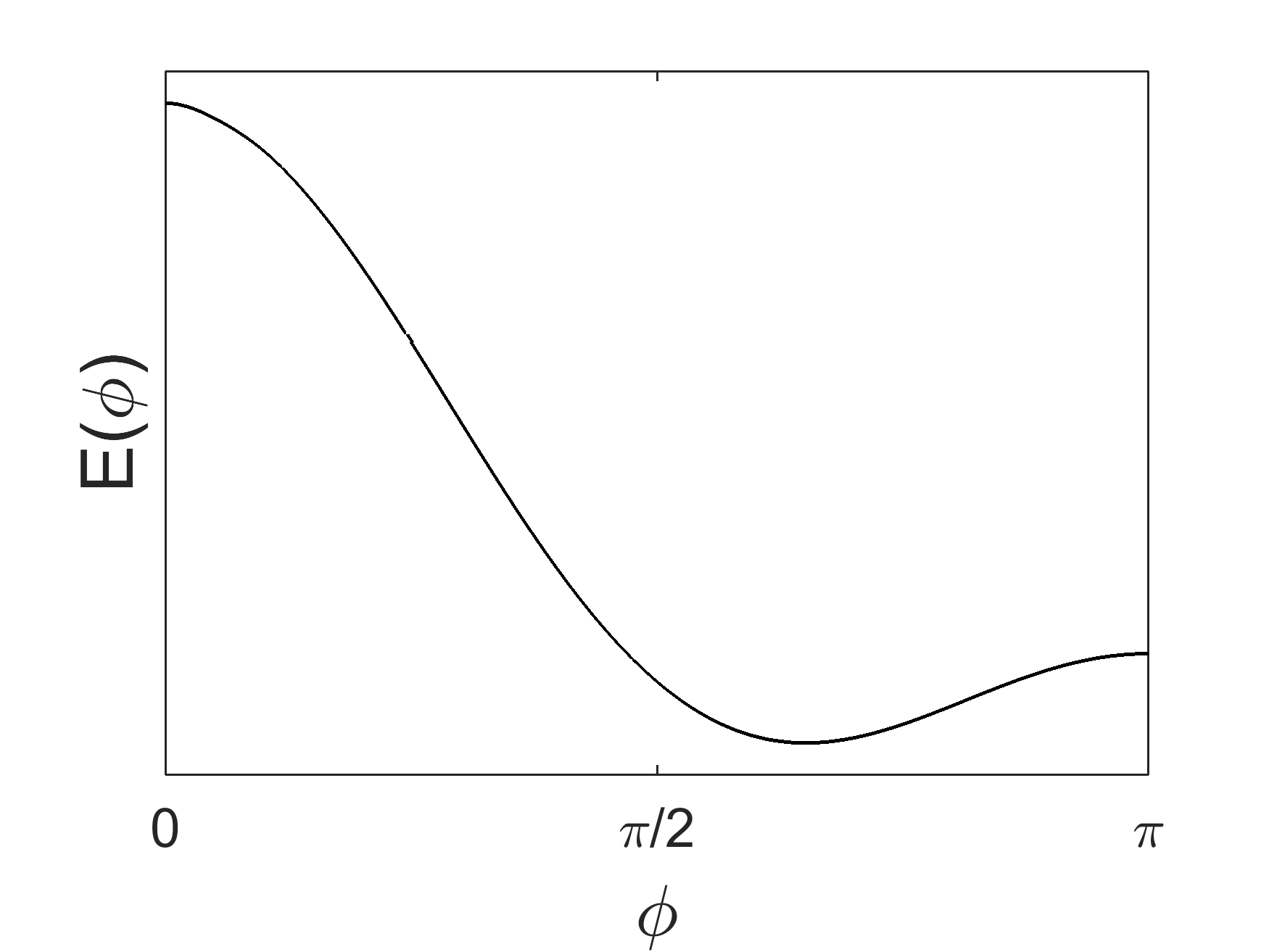}\label{fig:phijunction}}
\subfigure[\ ]{\includegraphics[width=0.48\linewidth]{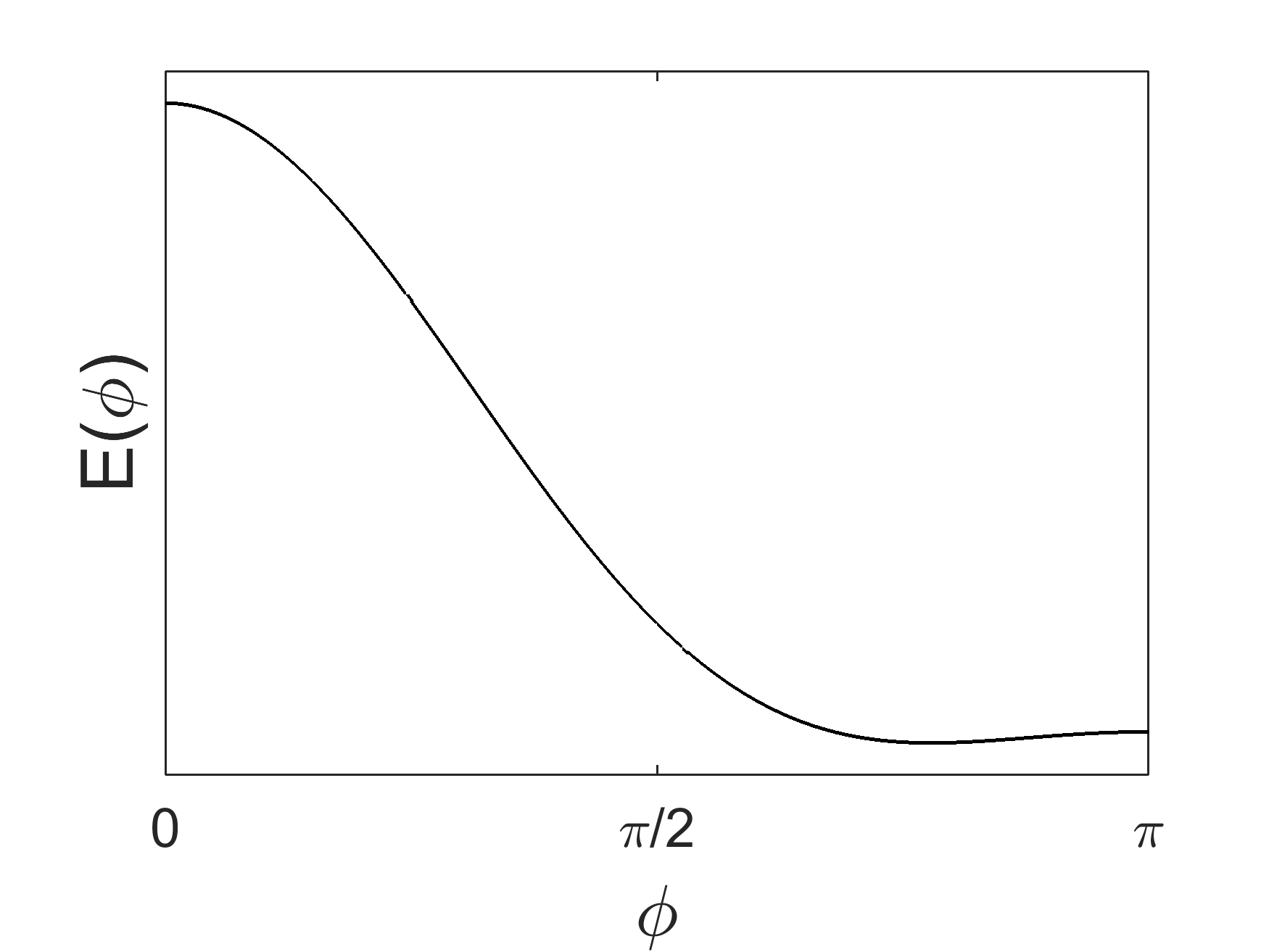}}
\subfigure[\ ]{\includegraphics[width=0.48\linewidth]{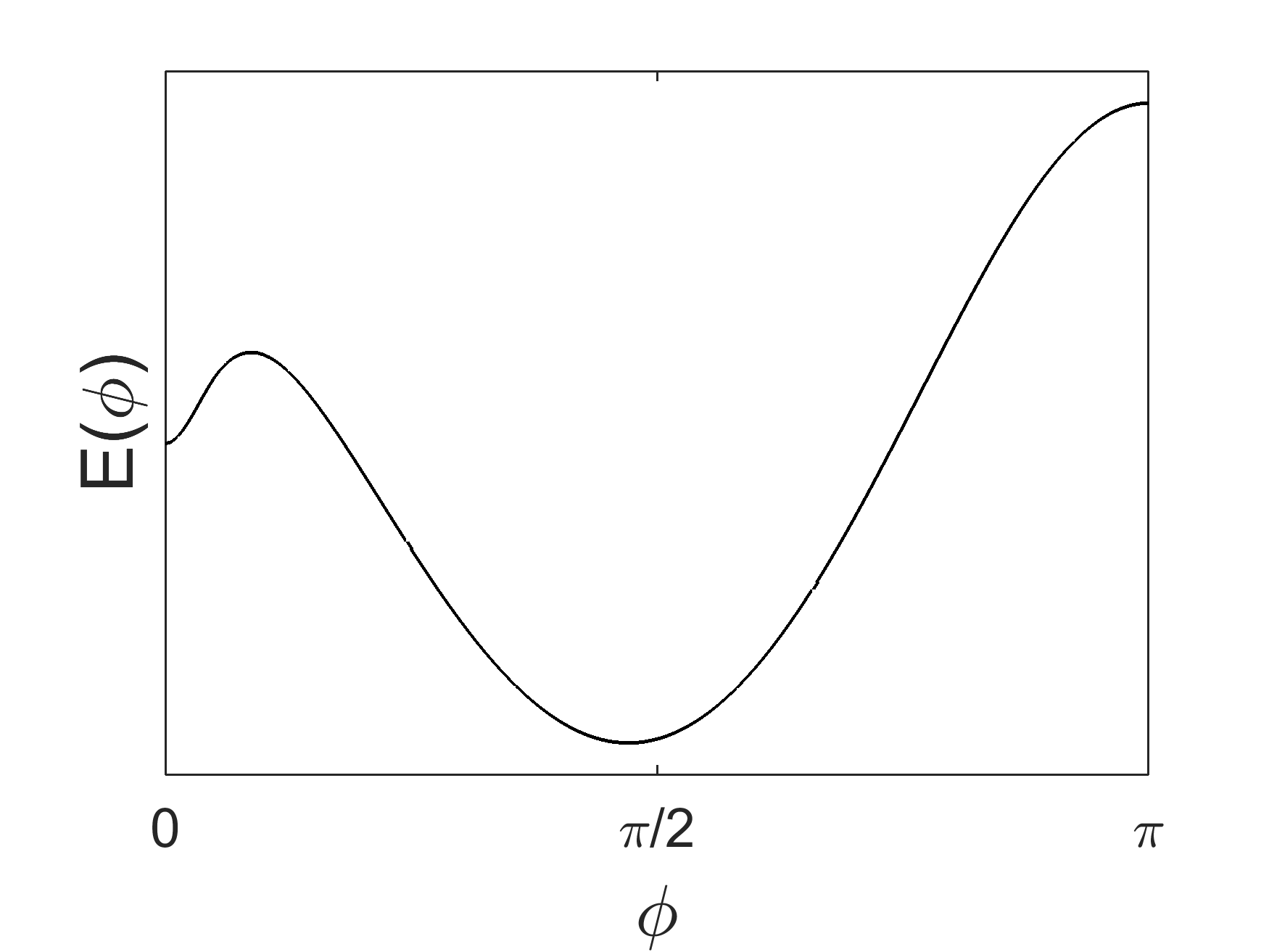}\label{fig:doubleminimum}}
\subfigure[\ ]{\includegraphics[width=0.48\linewidth]{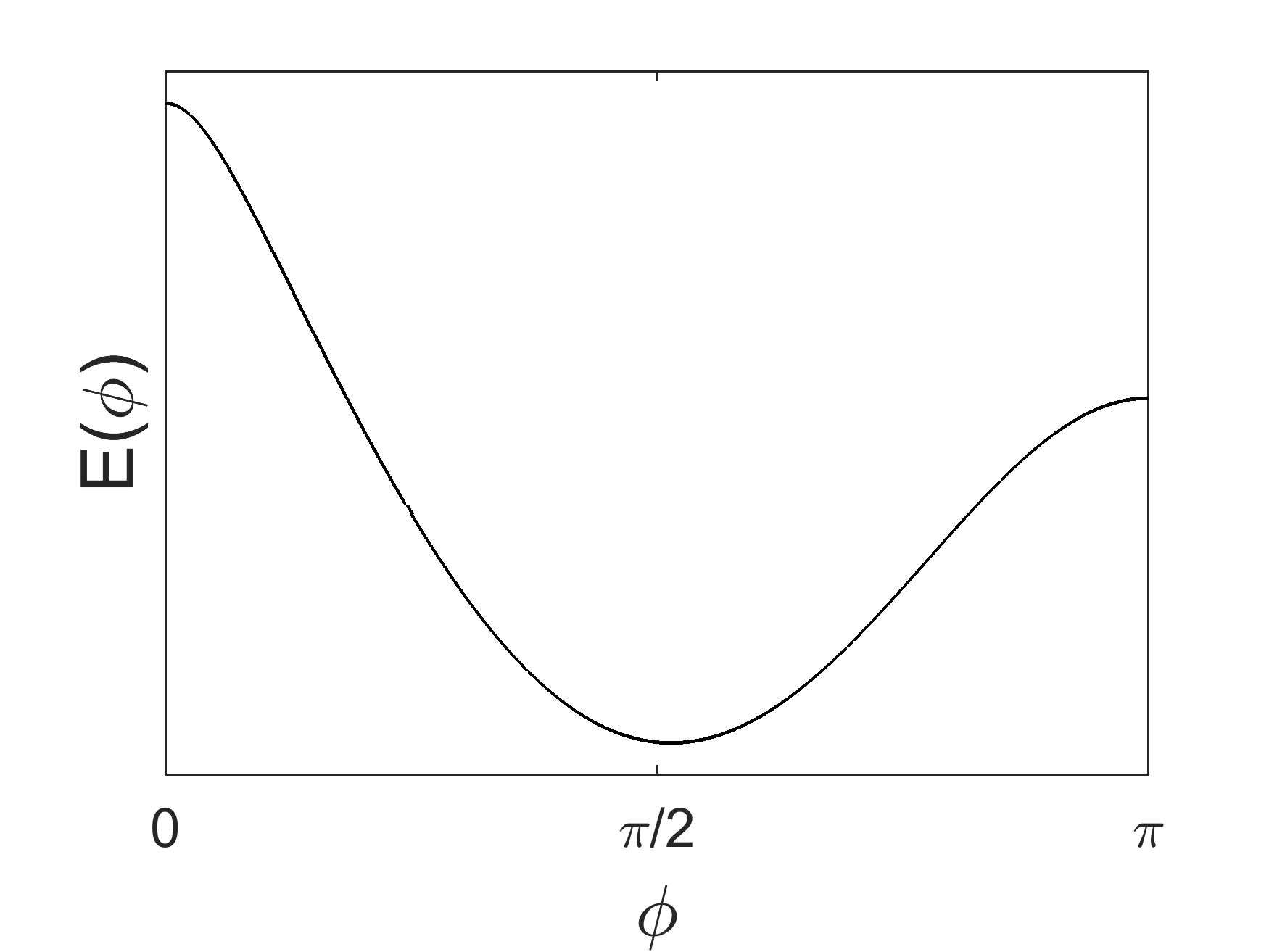}}
\caption{\label{fig:fig1} LEFT: Examples of energy vs.~phase behavior calculated self-consistently for (a) a 0-Junction and (c) a Double Minimum Junction. RIGHT: Panels (b) and (d) show the energy vs.~phase behavior obtained from non self-consistent calculations for the same parameters used in (a) and (c), respectively.}
\end{figure}

We have performed junction simulations on a 20 by 20 lattice for each of the $s$-wave and the $s_\pm$ superconductors, allowing the order parameters of the system to be determined self-consistently on up to 5 lattice sites from the contact. In order to elucidate the effects of the self-consistent determination of the order parameter, we also perform non-self-consistent energy and current calculations in which the order parameters are fixed.

All of the energy relations found in our model are $2\pi$ periodic and inversion symmetric ($E(\phi) = E(-\phi)$). We therefore present the energy-phase relation in the $[0,\pi]$ interval. For the studied parameter space, we find four types of junctions: a) 0-junctions where the energy is minimized when the phase difference between the order parameter of the $s$-wave superconductor and the hole pockets is 0 (corresponding to a phase difference of $\pi$ between the $s$-wave and the electron pockets), b) $\pi$-junctions, where the energy is minimized for phase difference $\pi$, c) $\phi$-junctions, see Fig.~\ref{fig:phijunction}, where the energy is minimized for a phase value $\phi$, with $0<\phi<\pi$ and d) double minimum junctions, see Fig.~\ref{fig:doubleminimum}, which present two minima in the $[0,\pi]$ interval, a local minimum at 0 and a global minimum at $0<\phi \leq \pi$. In the following text we use the term $\pi$-junction in the sense that the junction energy is minimized for a phase difference of $\pi$ between the $s$-wave and the $s\pm$-wave order parameter of the electron pocket, as described above.

\subsubsection*{Phase diagram}

\begin{figure}
\centering
\subfigure[\ ]{
\includegraphics[width=0.48\linewidth]{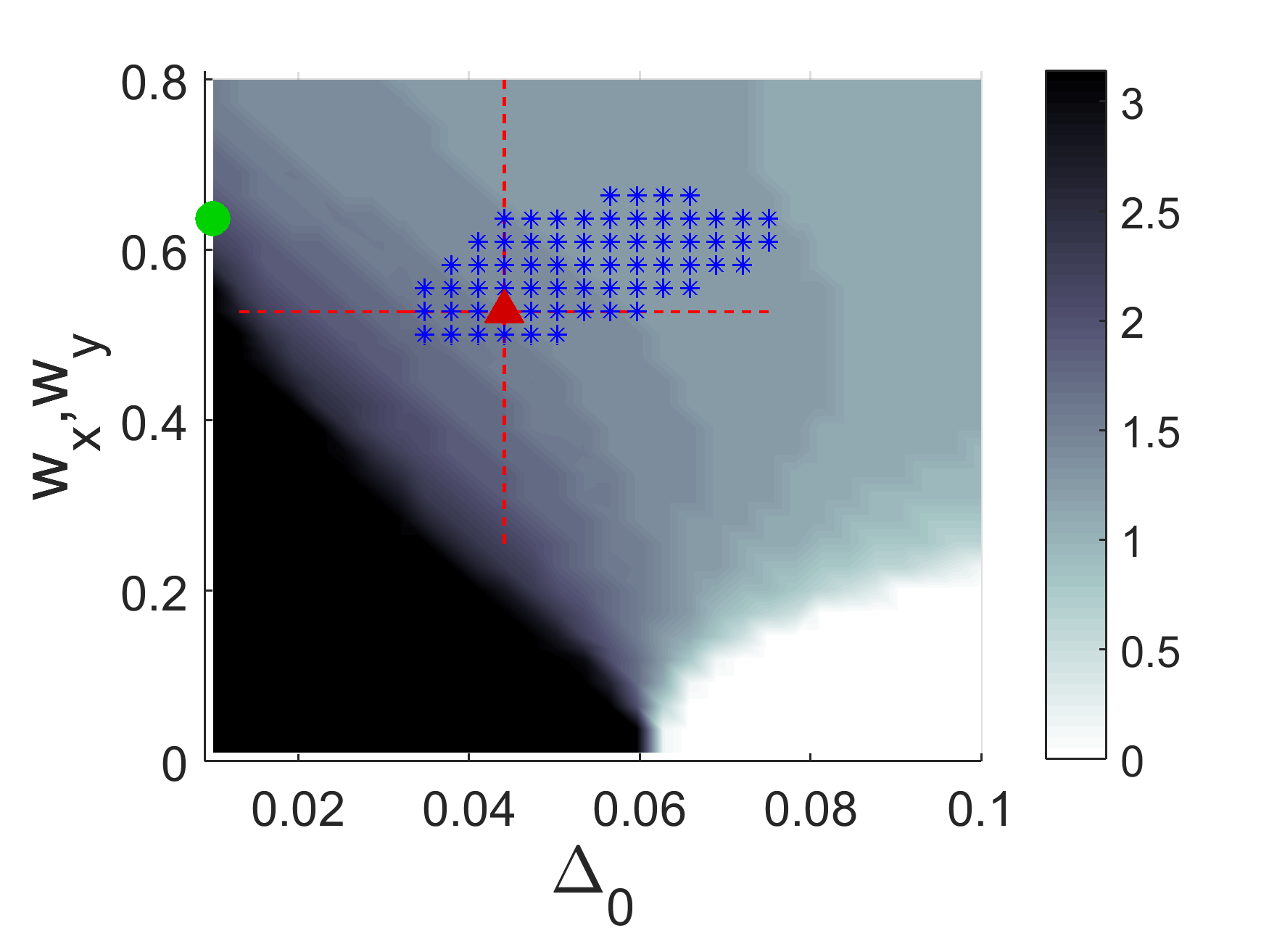}}
\subfigure[\ ]{
\includegraphics[width=0.48\linewidth]{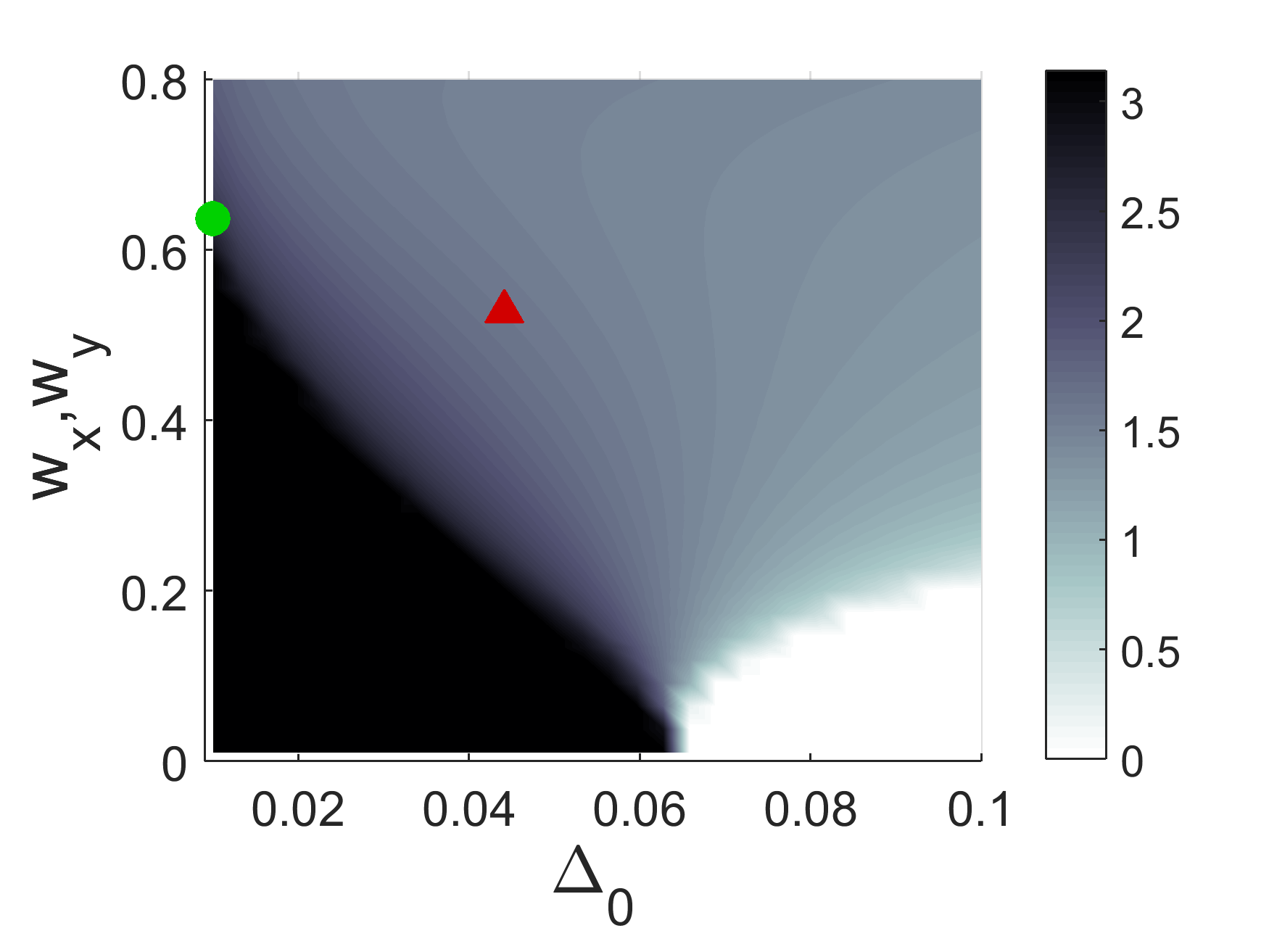}}
\caption{\label{fig:sc13} Value of the phase difference that minimizes the energy of an $s$-$s_\pm$ junction at zero temperature for $\mu_0=-1.3$ when (a) the order parameter is solved self-consistently for both superconductors close to the contact, and (b) the order parameter is constant on both sides. The areas marked by blue symbols represent a state with a double minimum energy/phase relation, where the global minimum is at $0<\phi\leq\pi$ while the local minimum is at 0 phase difference. All the parameters are given in units of $|t_1|$. The green circle corresponds to the parameters used in Fig.~\ref{fig:fig1}(a) and Fig.~\ref{fig:fig1}(b), and the red triangle to those of Fig.~\ref{fig:fig1}(c) and Fig.~\ref{fig:fig1}(d).} 
\end{figure}

\begin{figure}
\centering
\subfigure[\ $\Delta_0 = 0.06$]{
\includegraphics[width=0.48\linewidth]{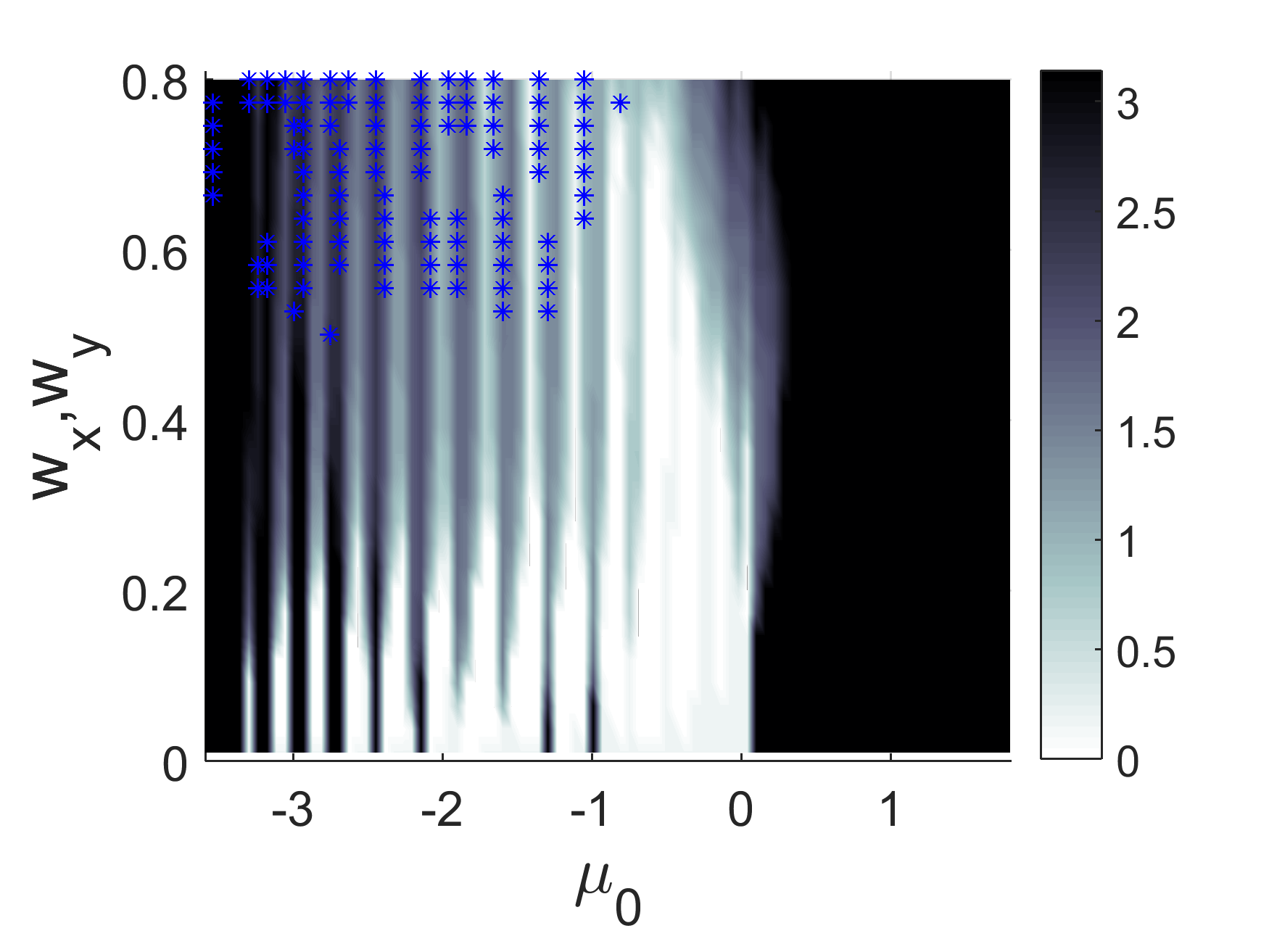}}
\subfigure[\ $\Delta_0 = 0.08$]{
\includegraphics[width=0.48\linewidth]{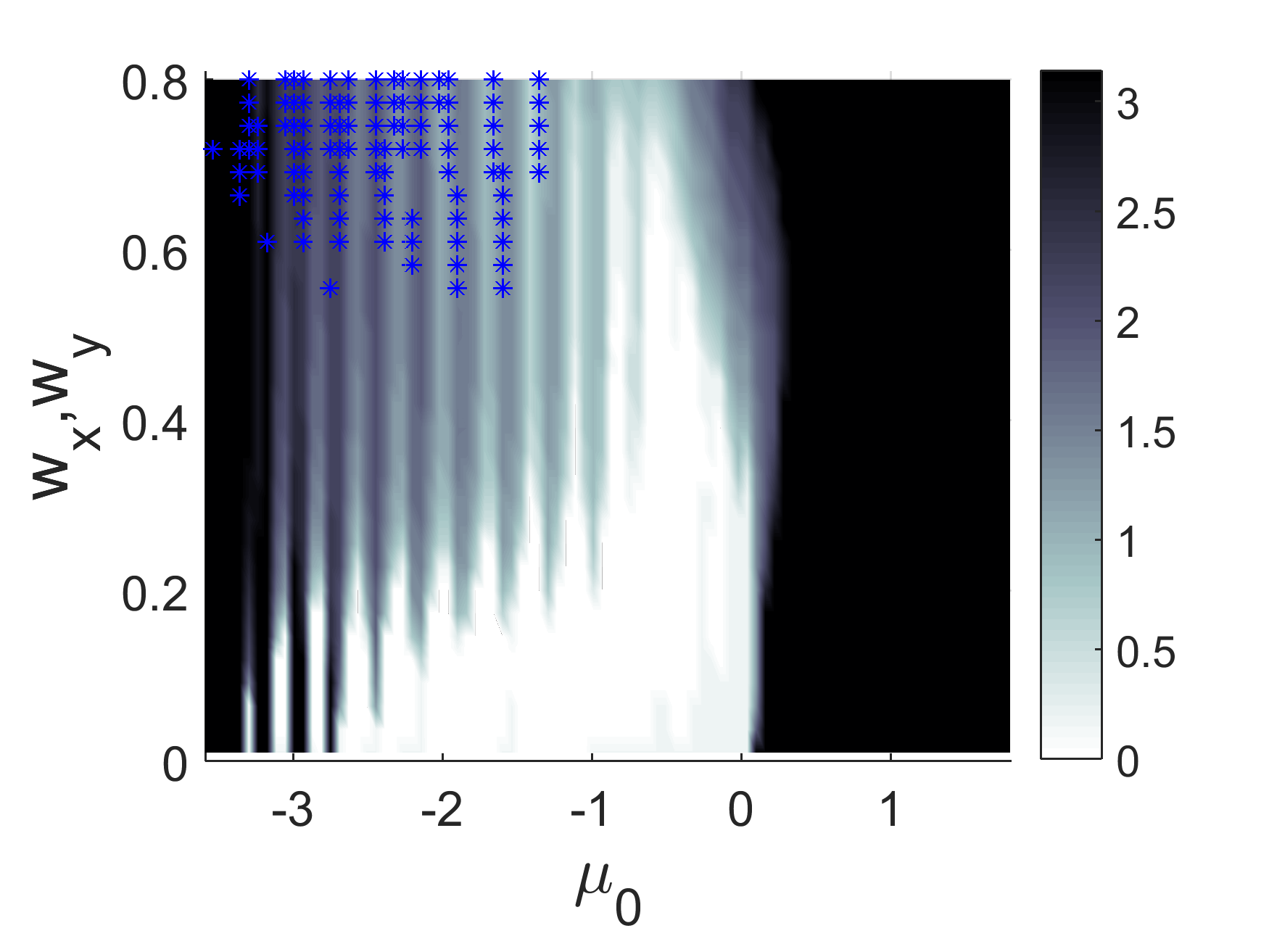}}
\subfigure[]{
\includegraphics[width=0.5\linewidth]{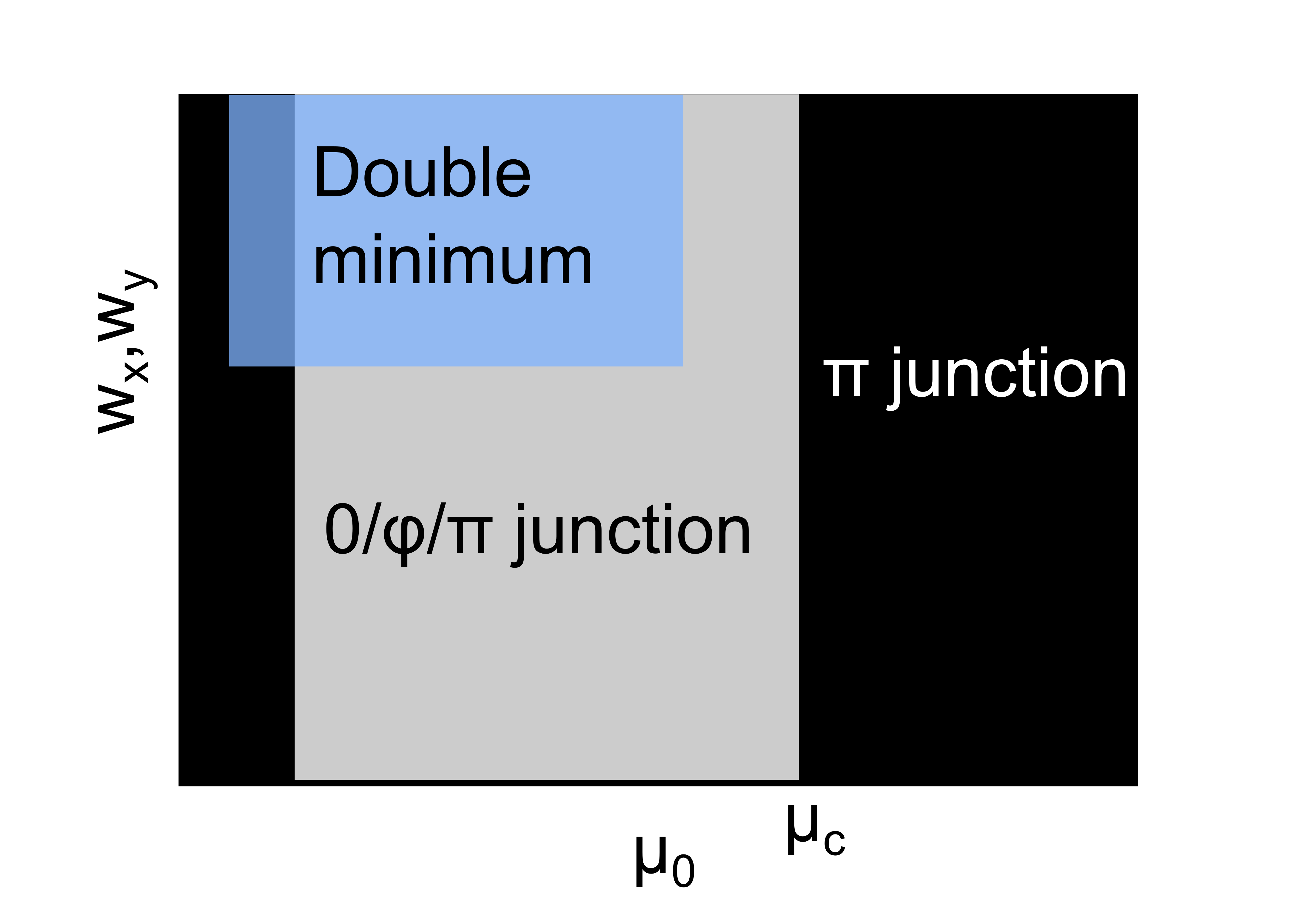}}
\caption{\label{fig:tvsmu} The panels (a) and (b) show the phase diagram by plotting the phase difference which minimizes the junction energy in color as a function of the tunneling amplitudes, $w_x=w_y$ and the chemical potential of the $s$-wave superconductor, $\mu_0$. White regions correspond to a phase difference of $0$ between the $s$-wave superconductor and the electron pocket order parameter while black color corresponds to a phase difference of $\pi$. The gray areas represent intermediate phase difference values and the blue symbols indicate areas with a double minimum energy/phase relation. Panel (c) is a schematic plot of the phase diagram. In the black areas the energy of the system is minimized by a phase difference of $\pi$, while in the gray area there is a close competition between the 0, $\phi$ and $\pi$ junction phases. In the area marked by blue there is a possibility of finding an additional minimum at $0$ phase difference.}
\end{figure}

\begin{figure}
\centering
\subfigure[\ \label{fig:orbital1}]{
\includegraphics[width=0.5\linewidth]{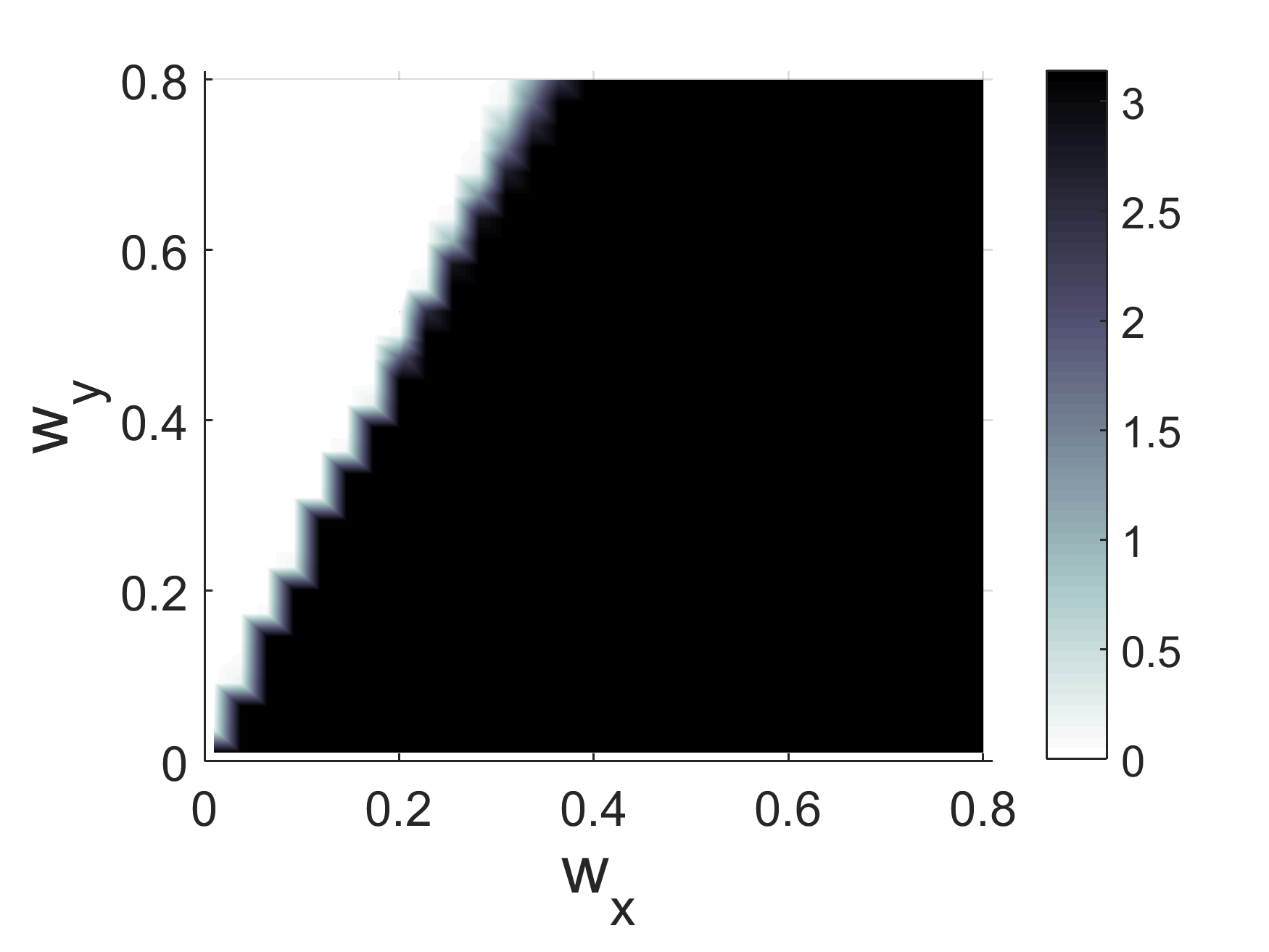}}
\subfigure[\ \label{fig:FS1}]{
\includegraphics[width=0.37\linewidth]{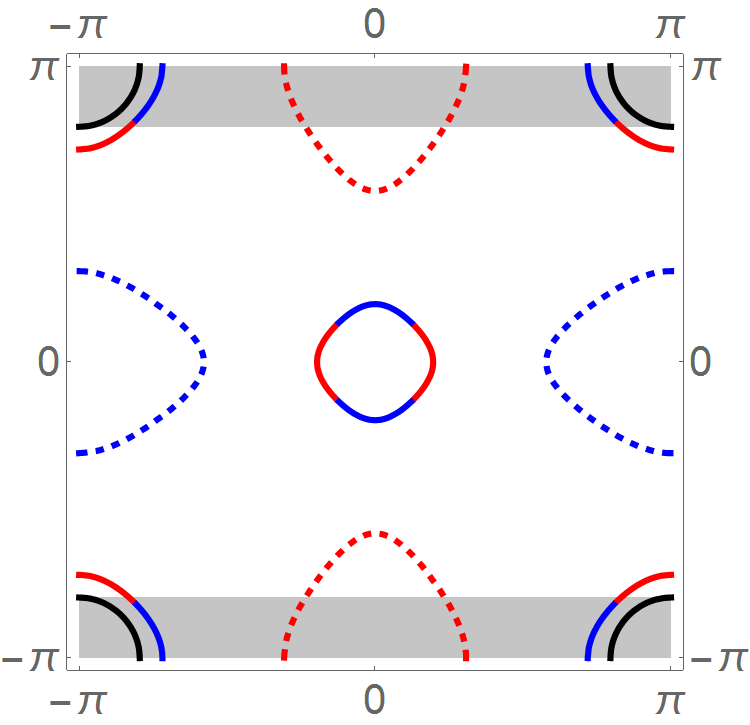}}
\subfigure[\ \label{fig:orbital2}]{
\includegraphics[width=0.5\linewidth]{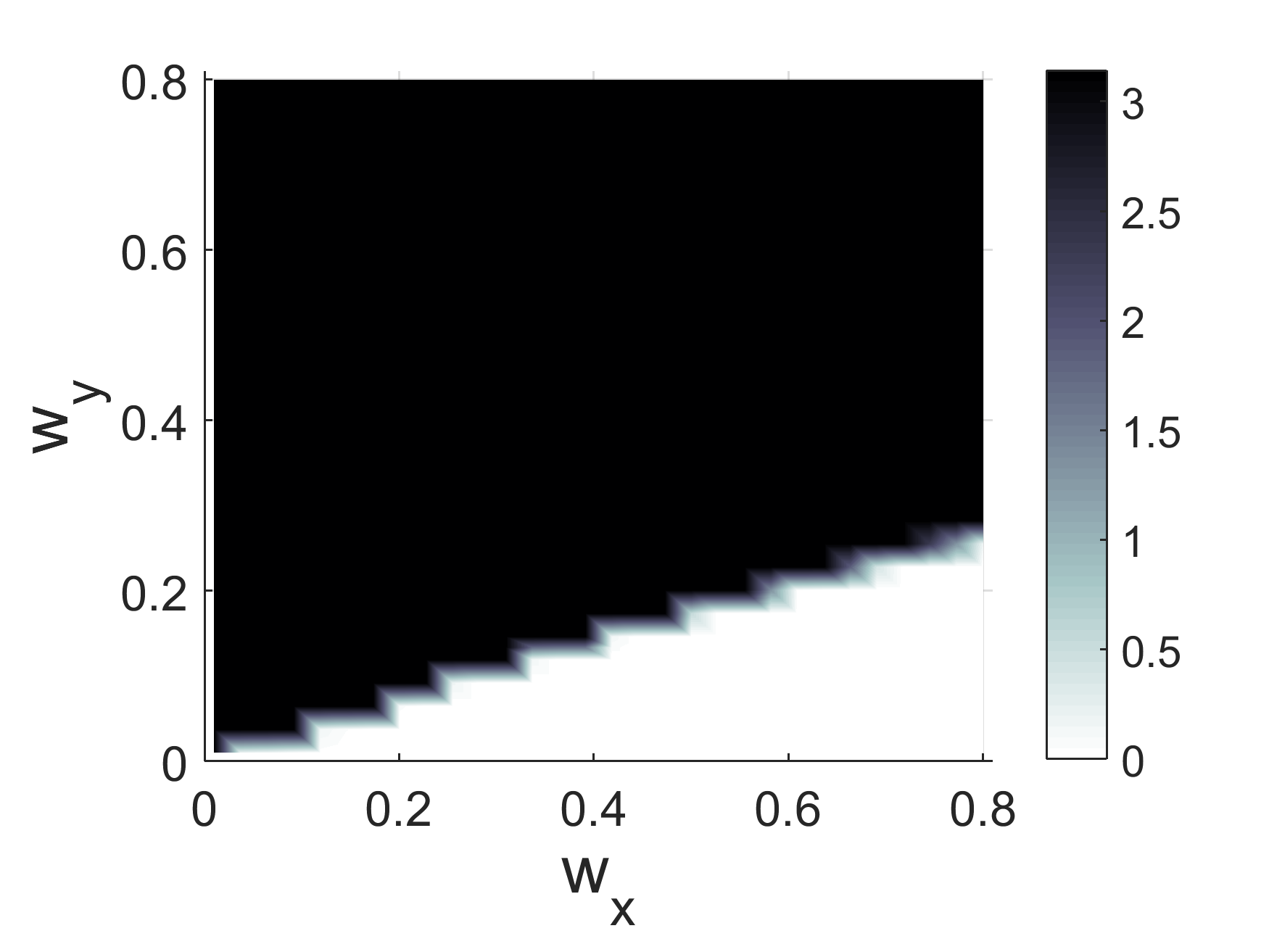}}
\subfigure[\ \label{fig:FS2}]{
\includegraphics[width=0.37\linewidth]{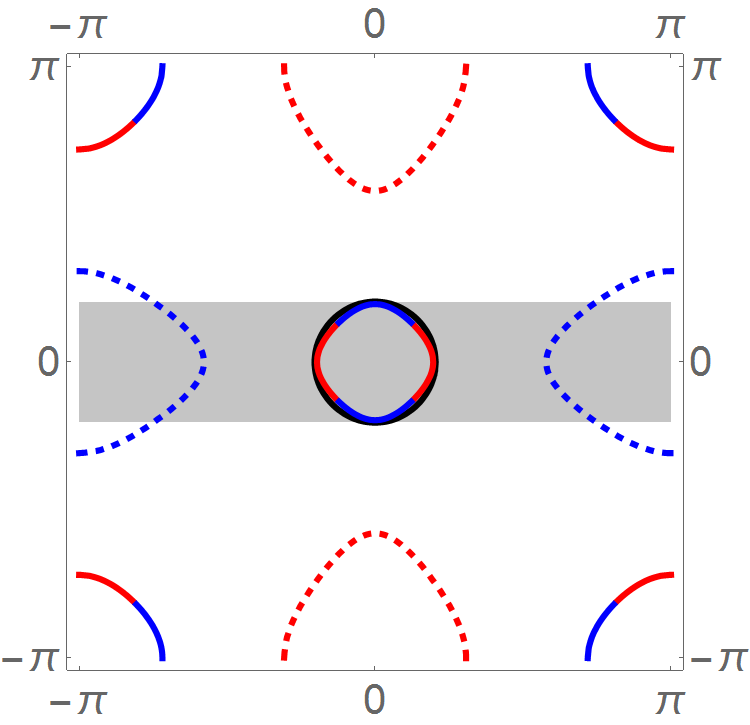}}
\subfigure[\ \label{fig:orbital3}]{
\includegraphics[width=0.5\linewidth]{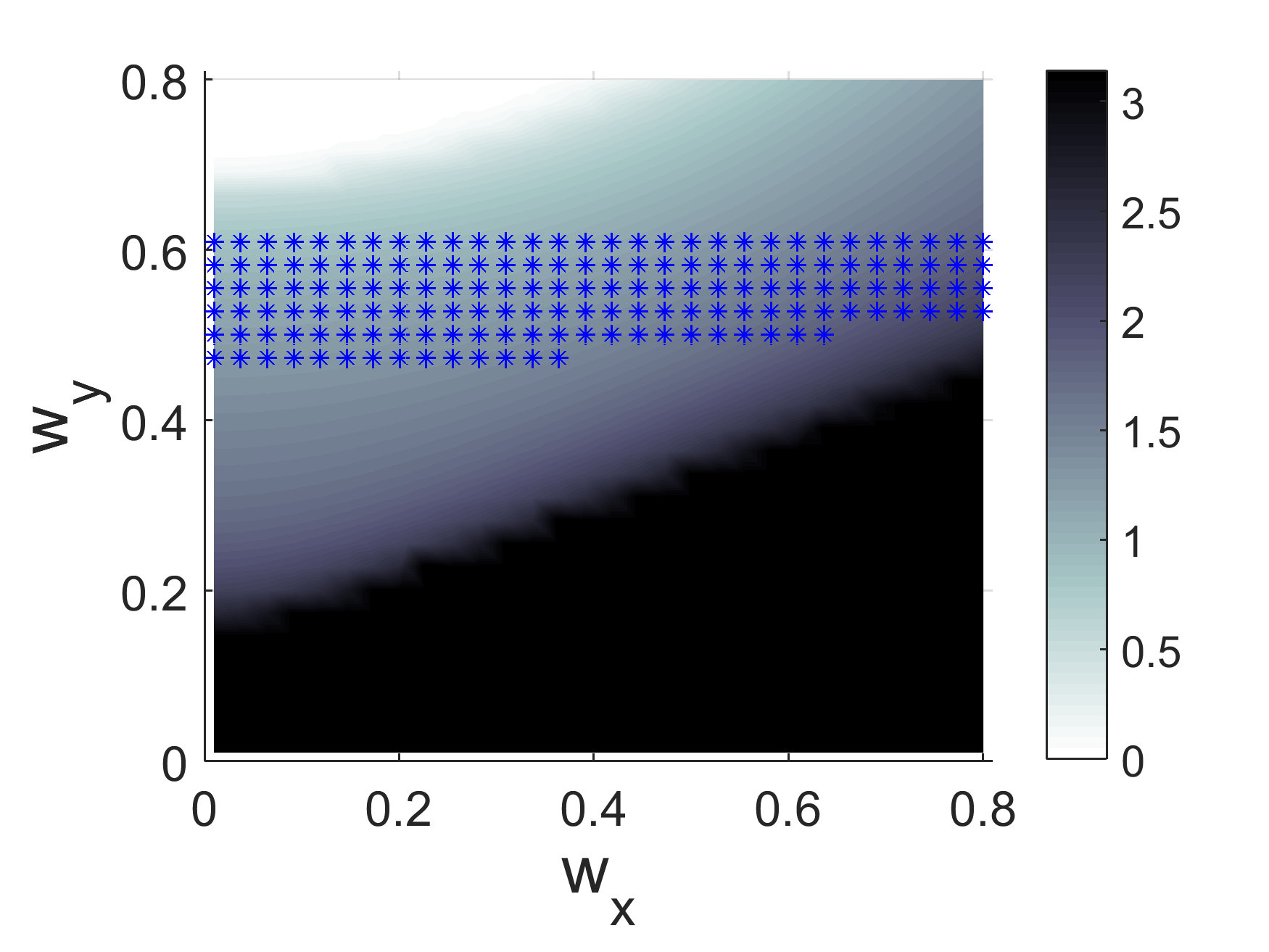}}
\subfigure[\ \label{fig:FS3}]{
\includegraphics[width=0.37\linewidth]{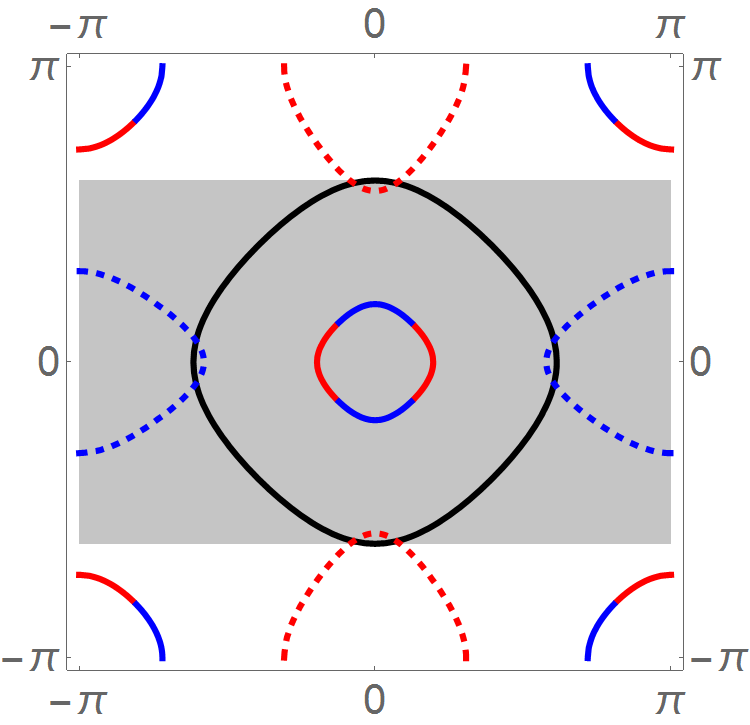}}
\caption{\label{fig:orbital} LEFT: The phase difference of the energy minimum as a function of $w_x$ and $w_y$ for (a) $\Delta_0 = 0.04$, $\mu_0=3.6$ and $U=2.33$, (c) $\Delta_0 = 0.04$, $\mu_0=-3.6$ and $U=2.33$ and (e) $\Delta_0 = 0.04$, $\mu_0=-1.3$ and $U=1.45$. RIGHT: The areas marked by blue symbols represent a state with a double minimum energy/phase relation. Fermi surface of the two superconductors in the extended Brillouin zone centered around the $\Gamma$ point for (b) $\mu_0=3.6$, (d) $\mu_0=-3.6$ and (f) $\mu_0=-1.3$. For the $s_\pm$ superconductor, the hole pockets marked by straight lines, and the electron pockets with dashed lines. The red/blue coloring indicates the portions of the Fermi surface whose main contribution comes from the $d_{xz}$/$d_{yz}$ orbital. The Fermi surface of the $s$-wave system is shown with a solid black line. The shading marks the values of $k_y$ for which there are $s$-wave Fermi surface states. Note that since the system is only invariant to translation in the $y$-direction $k_y$ is the only relevant momentum. All the parameters are given in units of $|t_1|$.}
\end{figure}

Our model contains parameters that characterize the properties of both superconductors and the contact between them. In order to study some representative phase diagrams we fix the bulk properties of the $s_\pm$ superconductor, and focus on varying the properties of the contact and the $s$-wave superconductor. Following Ref.~[\onlinecite{2bandmodel}] we set: $t_1 = -|t_1|$, $t_2=1.3|t_1|$ and $t_3=t_4=-0.85|t_1|$. For this choice of parameters, half filling corresponds to $\mu = 1.54|t_1|$. Since doping is a common way to tune the superconducting phase in FeSCs, we choose to work away from half-filling and set $\mu = 1.805|t_1|$. This value of $\mu$ corresponds to a doping of $0.18$ electrons per Fe site. Following the measurements from Refs.~[\onlinecite{Gap9.3meV,Gap7meV}] and the estimates for $|t_1|$ found in Ref. [\onlinecite{overlap}], we set the bulk pairing amplitudes $\Delta_x(n,n+1) = \Delta_y(n,n+1)=0.08|t_1|$. In the subsequent text we work in units such that $|t_1|=1$. On the $s$-wave superconductor, we fix $t_0= |t_1|$ to get a similar band width on the two sides of the junction. The other parameters of the $s$-wave superconductor, i.e. the chemical potential $\mu_0$ and the bulk pairing amplitude $\Delta_0$, as well as the contact parameters are varied. Sample phase diagrams are shown in Fig.~\ref{fig:sc13}, \ref{fig:tvsmu} and \ref{fig:orbital}.

Fig.~\ref{fig:sc13} demonstrates the importance of treating the order parameter close to the junction self consistently. The phase of junctions with a double minimum in the energy-phase relation only appears when the model is treated self consistently. We also find that the phase boundaries between the $\pi$-junction, $0$-junction and the $\phi$-junction are shifted in the self-consistent treatment as compared with the non-self-consistent one.

Observing the energy/phase relations, $E(\phi)$ we see that the location of the global minimum changes continuously as the model parameters are varied. If we start in a phase where the minimum is at $\phi=0$ and change the parameters the minimum changes continuously until it reaches $\pi$. Hence, the $\phi$-junction is always between the $0$-junction and the $\pi$-junction. When the tunneling parameters of the contact $w_x,w_y$ are small, the transition between the $0$- and $\pi$-junction is very sharp and the $\phi$-junction phase occupies only a narrow sliver in parameter space. As the tunneling amplitudes are increased, the $\phi$-junction takes up a larger portion of parameter space. 

The parameter $\mu_0$, the chemical potential of the $s$-wave superconductor has a dramatic effect on the phase diagram. We can relate this to the overlap between the hole and electron pockets with the Fermi surface underlying the $s$-wave superconductor. This relation is further demonstrated in Fig.~\ref{fig:tvsmu}, where we slice the phase diagram along the line $w_x=w_y$ and a constant order parameter $\Delta_0$. In all panels of Fig.~\ref{fig:tvsmu} there is a critical value of the chemical potential, $\mu_0$ such that for $\mu_0>\mu_c>0$, the energy of the junction is minimized when the phase difference is $\pi$ and for $\mu_0<\mu_c$, small changes in $\mu_0$ can lead to transitions from $0$ to $\pi$ minimum. This striking behavior with respect to the chemical potential, $\mu_0$, is also found when the order parameter is not solved self-consistently.

We can also observe in Fig.~\ref{fig:tvsmu} that for some values of $\mu_0$ and $\Delta_0$, transitions from $0$ to $\pi$ minimum can be driven by increasing the tunneling $w_x=w_y$. These transitions become more rare with increasing $\Delta_0$.

The final insight that can be gained from Fig.~\ref{fig:tvsmu} is that the double minimum behavior becomes more common as we increase $\Delta_0$, as well as the tunneling strength $w_x=w_y$.

The role of the tunneling parameters $w_x$ and $w_y$ is further explored in Fig.~\ref{fig:orbital}. The phase diagrams at zero temperature for $\Delta_0=0.04$ and three different values of the chemical potential of the $s$-wave side, $\mu_0=3.6$, $\mu_0=-3.6$, and $\mu_0=-1.3$, are shown in Fig.~\ref{fig:orbital1}, Fig.~\ref{fig:orbital2} and Fig.~\ref{fig:orbital3}, respectively.
The orbital composition of the Fermi surface for the $s_\pm$ superconductor together with the Fermi surface of the $s$-wave superconductor is shown in Fig.~\ref{fig:FS1} for $\mu_0=3.6$, in Fig.~\ref{fig:FS2} for $\mu_0=-3.6$ and in Fig.~\ref{fig:FS3} for $\mu_0=-1.3$. The contact preserves the momentum in the $y$-direction, hence fixing the value of $\mu_0$ will select the values of $k_y$ for which Cooper pairs can tunnel through the contact. For $\mu_0=3.6$, as shown in Fig.~\ref{fig:orbital1}, a transition from an energy minimum at $\pi$ to an energy minimum at 0 is driven by increasing the ratio $w_y/w_x$. It can be seen in Fig.~\ref{fig:FS1} that for this value of $\mu_0$ the pairs from the electron pockets that tunnel through the contact come from the $d_{xz}$ orbital while the pairs that tunnel from the hole pockets come mostly from the $d_{yz}$ orbital. The role of the parameters $w_x$ and $w_y$ for $\mu_0=-3.6$ in Fig.~\ref{fig:orbital2} is the opposite of the one in Fig.~\ref{fig:orbital1}, an energy minimum at 0 is obtained when the ratio $w_y/w_x$ is sufficiently small. The orbital composition of the pairs tunneling from the $s_\pm$ superconductor to the $s$ superconductor for $\mu_0=-3.6$ is shown in Fig.~\ref{fig:orbital2}. In this case, the electron-like pairs tunneling through the contact come from the $d_{yz}$, while the pairs coming from the hole pockets are evenly composed of $d_{xz}$ and $d_{yz}$ electrons.

For a large $s$-wave Fermi surface, the role of the parameters $w_x$ and $w_y$ is more difficult to understand. According to Fig.~\ref{fig:FS3}, the electron-like pairs tunneling through the contact mainly come from the $d_{yz}$ orbital, while the hole-like pairs are evenly composed of $d_{xz}$ and $d_{yz}$ electrons. Nonetheless, in Fig.~\ref{fig:orbital3} a $0$-junction phase is found for large enough $w_y$, which cannot solely be explained by looking at the orbital composition of the $s_\pm$ superconductor. The close competition between electron- and hole-like pairs for a large $s$-wave Fermi surface is also evident in Fig.~\ref{fig:orbital3} by the wide area of $\phi$-junction phase and the appearance of the double-minimum phase for large enough $w_y$. 

\subsubsection*{OP form}

Since the $s_\pm$ order parameter is defined on the lattice links we define an on-site order parameter in the $s_\pm$ superconductor for the purpose of visualization:
\begin{equation}
\Delta_{\alpha} (n) = \frac{1}{2} \left( \Delta_{\alpha} (n,n+1) + \Delta_{\alpha} (n+1,n)\right).
\label{eqn:onsite}
\end{equation}

\begin{figure}
\centering
\subfigure[\ \label{fig:opamp} $\phi=0$]{
\includegraphics[width=0.48\linewidth]{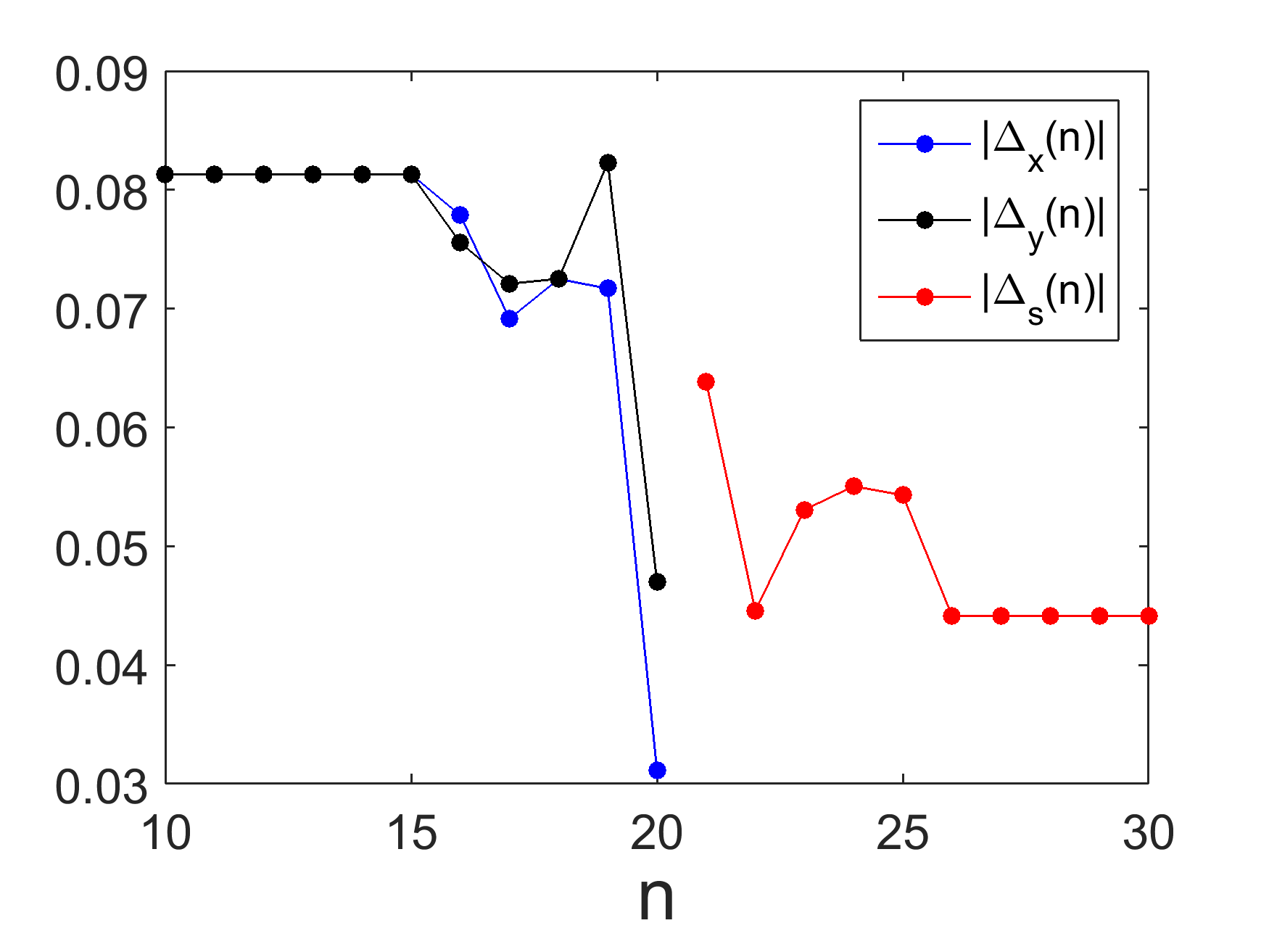}}
\subfigure[\ \label{fig:opphase} $\phi=0$]{
\includegraphics[width=0.48\linewidth]{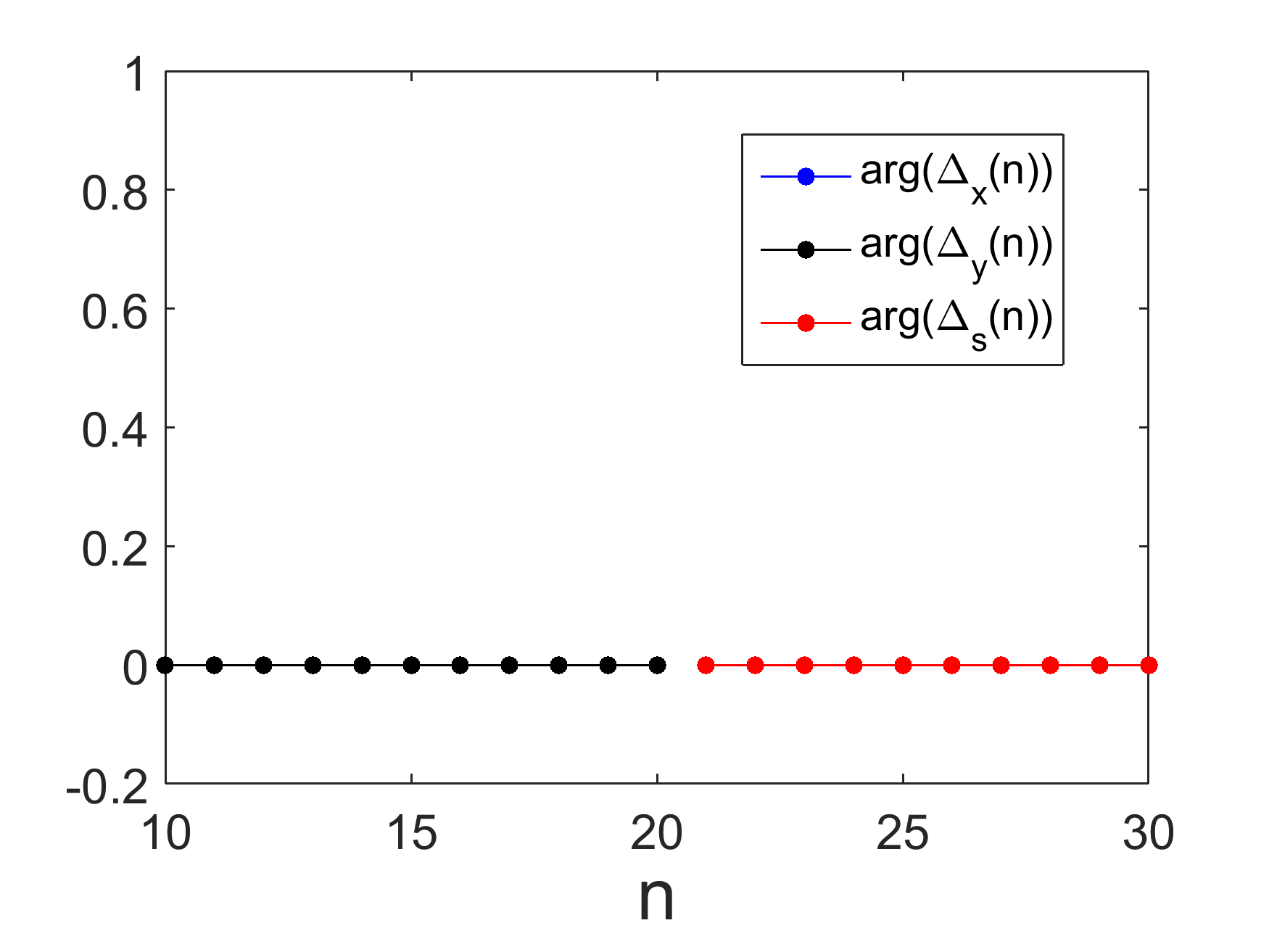}}
\subfigure[\ \label{fig:opamp2} $\phi=\pi/4$]{
\includegraphics[width=0.48\linewidth]{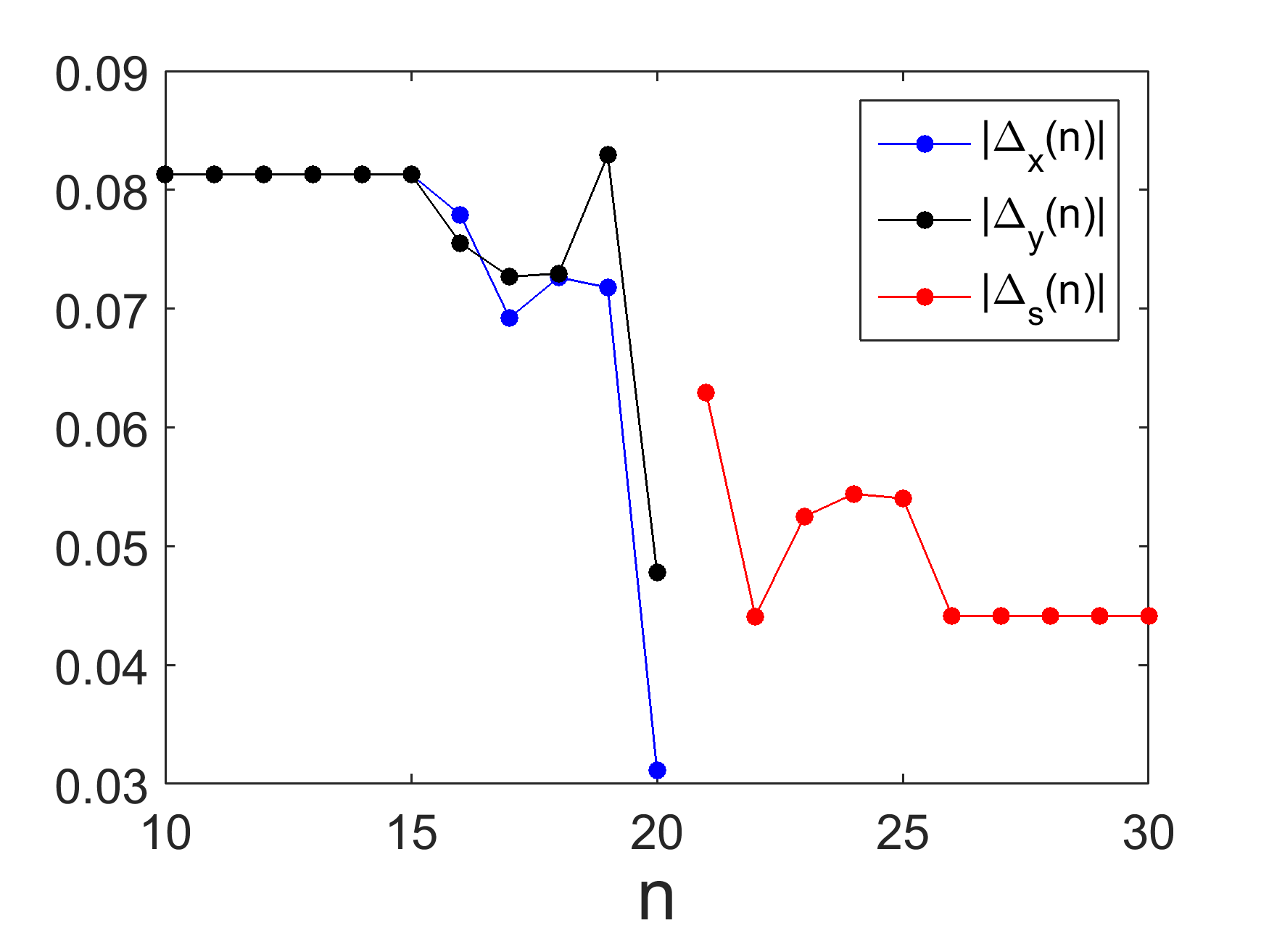}}
\subfigure[\ \label{fig:opphase2} $\phi=\pi/4$]{
\includegraphics[width=0.48\linewidth]{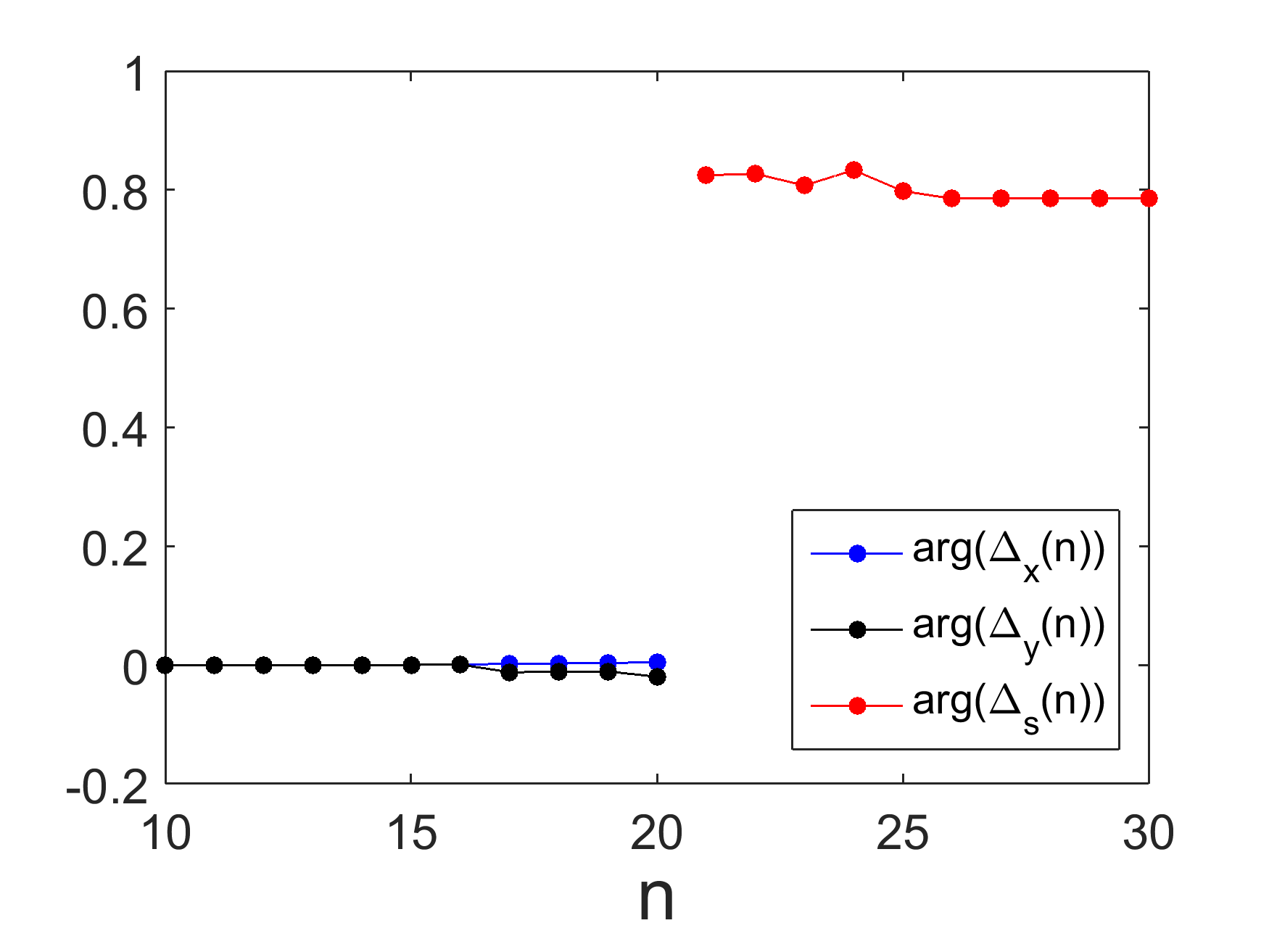}}
\caption{\label{fig:op} LEFT: Amplitude (in units of $|t_1|$) of the order parameters of the system for two values of the phase difference (a) $\phi=0$ and (c) $\phi= \pi/4$. RIGHT: Phase of the order parameters of the system for (b) $\phi=0$ and (d) $\phi= \pi/4$. The set of parameters used corresponds to the red triangle in Fig.~\ref{fig:sc13}.}
\end{figure}

In Fig.~\ref{fig:op} we look at the various order parameters as a function of their position with respect to the contact. We set the parameters of the system to the double minimum regime and plot the amplitudes in Fig.~\ref{fig:opamp} and Fig.~\ref{fig:opamp2}, and the order parameter phase in Fig.~\ref{fig:opphase} and Fig.~\ref{fig:opphase2} for two phase difference values, 0 and $\pi/4$. The amplitudes of the various order parameters of the system for 0 and $\pi/4$ phase difference, shown in Fig.~\ref{fig:opamp} and Fig.~\ref{fig:opamp2} respectively, presents only small differences. The variation of the phase of the order parameters from the bulk phase difference is imperceptible in Fig.~\ref{fig:opphase} and small in Fig.~\ref{fig:opphase2}. Despite having only small quantitative differences in the shape of the self-consistent solution for different phases, determining the order parameter self-consistently has important consequences for this choice of parameters, as it leads to the double minimum phase. 

In the different panels of Fig.~\ref{fig:op} we observe the presence of sharp oscillations. This kind of oscillations has been previously found in microscopic models of $s$-wave superconductors close to an insulating boundary and has been attributed to Friedel-like oscillations\citep{Friedel,Friedel2,Friedel3}. The period of the oscillations seen here is dependent on the chemical potential as expected.

\begin{figure}
\centering
\subfigure[\ \label{fig:optcut} ]{
\includegraphics[width=0.48\linewidth]{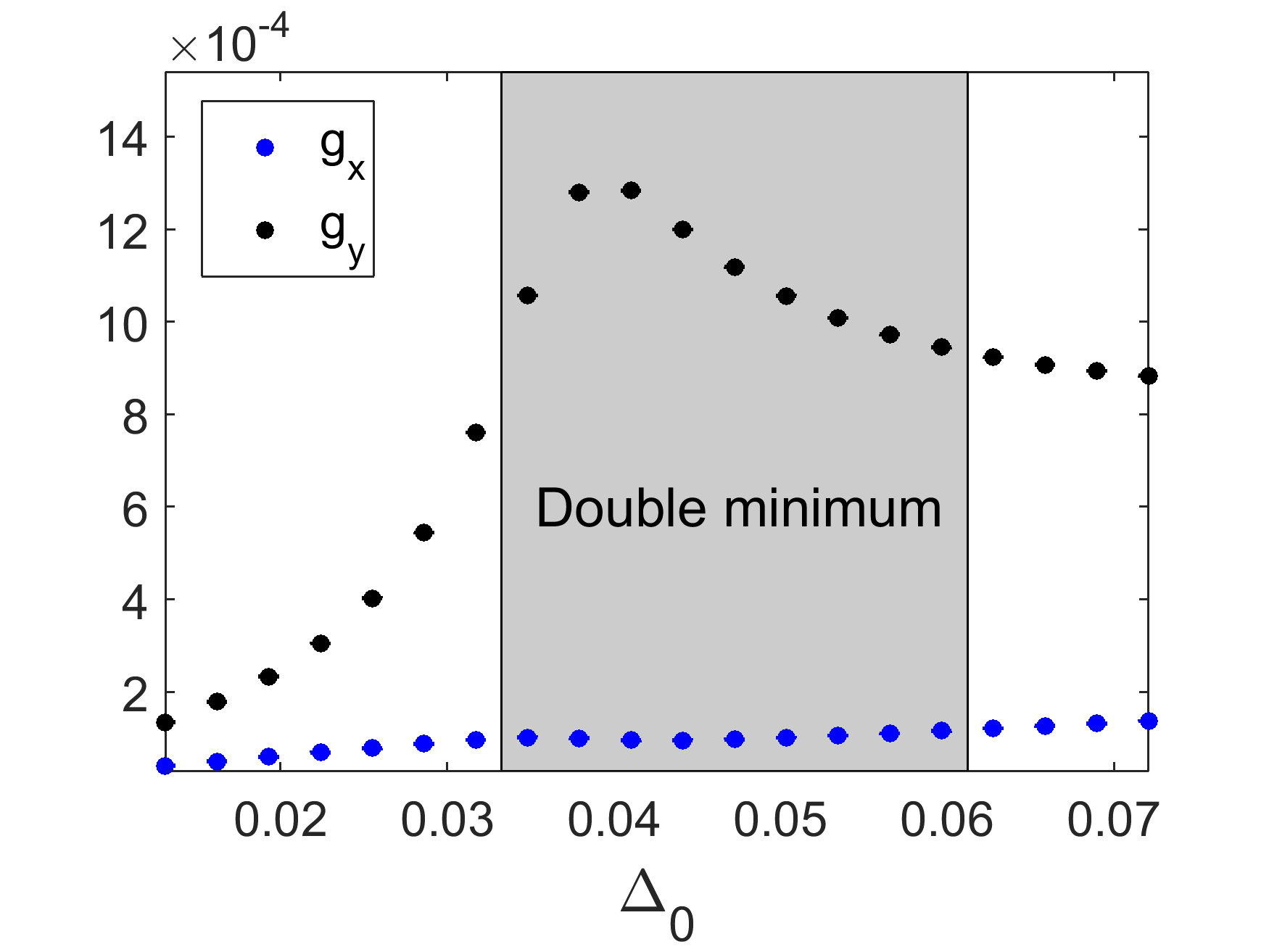}}
\subfigure[\ \label{fig:opdeltacut} ]{
\includegraphics[width=0.48\linewidth]{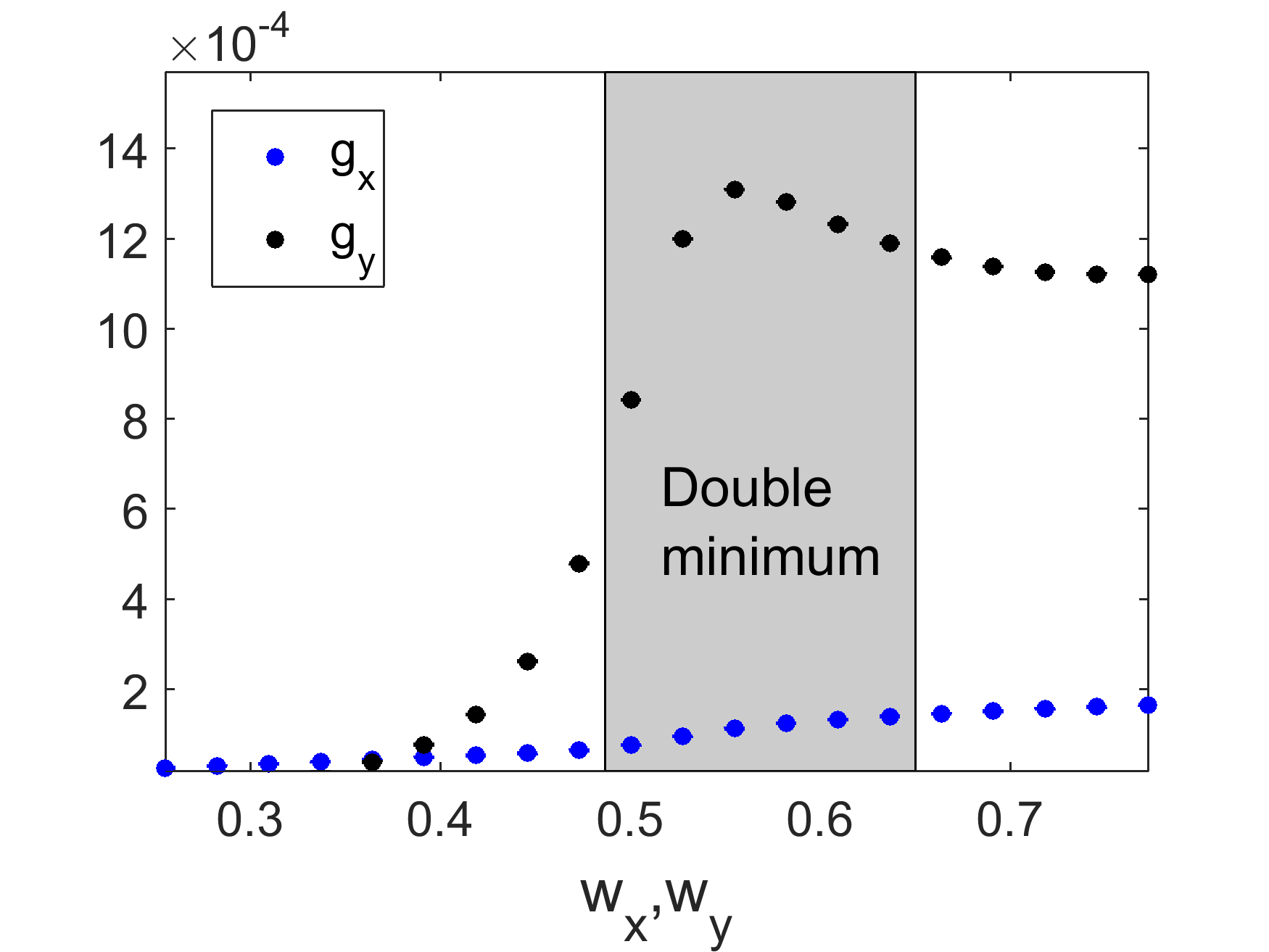}}
\caption{\label{fig:deltadependence} Values of $g_x$ and $g_y$ (in units of $|t_1|$) for the parameters given by the (a) horizontal and (b) vertical cuts marked with red on Fig.~\ref{fig:sc13}. The quantities $g_x$ and $g_y$, defined in Eqn.~\ref{eqn:gs}, measure the dependence of $\Delta_x$ and $\Delta_y$ with respect to the phase difference between the two superconductors.}
\end{figure}

In order to further explore the relation between the self-consistent determination of the order parameter and the double minimum phase, it is necessary to quantify the dependence of $\Delta_x$ and $\Delta_y$ on the phase difference $\phi$. To do this, we define the following functions:
\begin{equation}
\begin{split}
g_x =& \left\lvert \Delta_x(N_1,\phi=\pi) - \Delta_x(N_1,\phi=0) \right\rvert\\
g_y =& \left\lvert \Delta_y(N_1,\phi=\pi) - \Delta_y(N_1,\phi=0) \right\rvert .
\end{split}
\label{eqn:gs}
\end{equation}
The behavior of $g_x$ and $g_y$ for two different cuts of parameters around the double minimum phase is shown in Fig.~\ref{fig:deltadependence}. In both Fig.~\ref{fig:optcut} and Fig.~\ref{fig:opdeltacut}, $g_y$ is maximized in the double minimum regime, indicating that $\Delta_y$ has a greater dependence on the phase difference $\phi$ in the double minimum regime. On the other hand, the value of $g_x$ in Figs.~\ref{fig:optcut} and \ref{fig:opdeltacut} is an order of magnitude smaller than that of $g_y$, signaling a much lower dependence of $\Delta_x$ with $\phi$, and it increases slowly with increasing $w_x = w_y$ and $\Delta_0$. The stronger dependence of $\Delta_y$ (compared to $\Delta_x$) with $\phi$ is consistent with the greater role that $w_y$ (compared to $w_x$) has in driving the transition to the double minimum regime exhibited in Fig.~\ref{fig:orbital3}. The critical current of a Josephson junction increases when increasing the order parameter of the superconductors and the tunneling through the interface. Hence, the energy cost of the Josephson frustration is higher when $w_x$, $w_y$ or $\Delta_0$ increase, leading the to a stronger dependence of the order parameter on the phase difference as a mechanism to relieve this frustration. Accordingly, the double minimum state is more likely to appear when $w_x$, $w_y$ or $\Delta_0$ are large.

\subsection{Discussion}\label{sec:JJorigin}

The problem of Josephson tunneling between an $s_\pm$ superconductor and a single band $s$-wave superconductor has been considered previously using different approaches. These previous studies point at two possible scenarios: (i) the $s$-wave superconductor couples predominantly to either the electron or hole pockets or (ii) the couplings between the $s$-wave and the electron and hole pockets are comparable leading to Josephson frustration. The momentum space structure of the order parameter in our model is $\Delta_\alpha \cos(k_x)\cos(k_y)$, hence in the hole pockets the order parameter phase is the same phase as $\Delta_\alpha$ while in the electron pockets the order parameter phase is shifted by $\pi$. Hence, we can interpret the $0$-junction as a scenario in which the $s$-wave is interacting primarily with the hole pockets, the $\pi$-junction as one where the $s$-wave is interacting primarily with the electron pockets, and the other two cases ($\phi$-junction and a double minimum junction) as resulting from competing interaction of the $s$-wave with electron an hole pockets.

In the current study, we find that the global minimum is at a phase difference of $\pi$, i.e. when the phase of the $s$-wave order parameter is equal to that of the electron-like pairs, for a large portion of the parameter space. This is consistent with the fact that we are working with an electron doped $s_\pm$ sample. Hence, the $s_\pm$ superconductor has more pairs that come from electron pocket states rather than from hole pocket states. Upon tuning of model parameters, the preferred phase difference may shift to $0$. In both these cases the system minimizes its energy by matching the OP phase of the $s$-wave superconductor with either the hole or electron pockets. This tendency is strongly dependent on the overlap between Fermi surface states on both sides of the Junction.
Figure \ref{fig:orbital} shows the orbital composition of different parts of the $s_\pm$ Fermi surface, at two different chemical potentials on the $s$-wave side and with varying tunneling amplitudes. It should be noted that while the hole pockets are composed of both the $x$ and $y$ orbitals, each electron pocket is dominated by one orbital. Moreover, since the contact conserves momentum only in the $y$-direction, the chemical potential of the $s$-wave part selects which parts of the $s_\pm$ Fermi surface participates in the tunneling. Depending on the value of $\mu_0$, the electron-like and hole-like pairs involved in the tunneling process have a different orbital composition. Thus, the ratio of $w_y/w_x$ at which the junction switches from the 0-junction to the $\pi$-junction phase depends on the value of $\mu_0$ and the geometry of the junction.

Our phase diagrams suggest that a transition between a 0-junction and a $\pi$-junction does not occur directly. Instead, an intermediate $\phi$-junction or double minimum phase appears (see Fig.~\ref{fig:phijunction} and ~\ref{fig:doubleminimum}).
Despite being a regime commonly found in theoretical models, there is no consensus on the mechanism behind these junction states. While Ref.~[\onlinecite{Microscopic2}] stresses that the extent of the $\phi$-junction is related to the second Josephson harmonic, this phase is also obtained within a Ginzburg-Landau mechanism which only considers first order terms in the Josephson coupling. In this case, the mechanism that leads to the $\phi$/double minimum phase is the possibility of twisting the relative phase between the hole-like and electron-like pairs.

In this work, we developed a model that considers both higher order Josephson terms and a self-consistently determined order parameter. Our results suggest that the higher order Josephson harmonics play a more important role in the realization of the $\phi$ state. This can be demonstrated by comparing the results of a self-consistent order parameter determination with the non-self consistent solutions. In the non-self-consistent case, the hole and electron pocket order parameters are given by the form of Eq.~\ref{eqn:spmParams}. Unlike in the self-consistent treatment, their amplitudes are related and their phase difference is always $\pi$. Comparing the two cases we see that the main difference is the appearance of the double minimum phase (Fig.~\ref{fig:sc13}). A secondary effect is a slight shift in the boundaries of the $\phi$-junction state in the phase diagram. We therefore conclude that the main reason for the appearance of the $\phi$-junction is the inclusion of higher harmonics in the Josephson tunneling. This is also supported by the fact that the tendency to develop a $\phi$-junction is increased when the tunneling amplitude across the junction is increased.

To make this point clearer, let us look into the first two terms in the Josephson coupling. It has been pointed out in Ref.~[\onlinecite{secondharmonicballistic}] that if the coupling between the hole-like pairs and the $s$-wave superconductor is similar to the coupling between the electron-like pairs and the $s$-wave superconductor, the first order terms on the Josephson energy tend to cancel each other, increasing the importance of the next order terms. This can be shown by writing the Josephson energy of the junction as $E= E_h +E_e$, where $E_{h(e)}$ is the Josephson energy associated to the coupling between the $s$-wave and the hole(electron) pockets. Then we have:
\begin{equation}
E_{h(e)} = E_{e(h)}^{(1)} \cos \left(\phi_{h(e)} \right) + E_{h(e)}^{(2)} \cos \left(2 \phi_{h(e)} \right)+ ...,
\end{equation}
where $\phi_{h(e)}$ is the phase difference between the $s$-wave and the hole(electron) pairs. Since $\phi_e=\phi_h+\pi$, the energy of the contact is then
\begin{equation}
\begin{split}
E =&\left(E_h^{(1)}- E_{e}^{(1)}\right) \cos \left(\phi_{h} \right) + \\ &\left( E_h^{(2)}+E_e^{(2)} \right) \cos \left(2 \phi_{h} \right) + ...
\end{split}
\end{equation}
Hence, when $ E_h^{(1)} \approx E_{e}^{(1)}$, the first Josephson harmonic cancels , and the second order term cannot be neglected.

Several theoretical models of an $s_\pm$-$s$ junction can exhibit 0-junction, $\pi$-junction, or $\phi$-junction behavior\citep{quasiclassical,quasiclassical2,Microscopic2,multibandjunctions-sdw,secondharmonicballistic,TRBGL1,TRB2,Lagrangian,secondharmonicballisticprevious,
Microscopic2previous,wire,temperature,trilayerproposal,PhysRevB.66.214507,GLold,PhysRevB.91.214501}. On the other hand, a double minimum behavior has only been previously found in Ref.~[\onlinecite{metastablepijunction}] within a Ginzburg-Landau formalism. There are important differences between our results and those of Ref.~[\onlinecite{metastablepijunction}], namely, in the location of the two minima. The double minimum phase found in Ref.~[\onlinecite{metastablepijunction}] is characterized by one global minimum at 0 phase difference across the junction and a local minimum at $\pi$ phase difference. In our microscopic model, the `double minimum' junction behavior also occurs when there is a global minimum at some phase $\phi\neq 0,\pi$ and a local minimum at 0 phase difference.

It is interesting to note the differences and similarities between this work and the phenomenological Ginzburg-Landau treatment of Ref.~[\onlinecite{metastablepijunction}]. While the two models describe an interface between a single order parameter superconductor and a double order parameter superconductor the details are quite different. Most importantly, in the Ginzburg-Landau model the order parameters are defined on each band and are only coupled through their amplitudes (in a way that ensures their opposite sign is favored in the bulk). In our microscopic model, the order parameter is defined on the orbitals and due to inter-orbital hopping, the two bands are gapped. As a result, in the current work we can not easily control the gap magnitude on each band, nor can we directly control the effective coupling between the order parameters on the two bands. We should also note that while the Ginzburg-Landau model does not explicitly contain terms of higher harmonic Josephson tunneling, our model does. These differences make the comparison between the models difficult. However, we can speculate that the differences mentioned above are responsible for the different states found in the two models.

Overall, we find that for competing coupling between the electron and hole pockets and the $s$-wave it is important to consider higher order processes, such as higher order Josephson harmonics and the effect of the contact on the order parameters of the system. At low tunneling strength, the contact phase diagram is dominated by the $0$ -junction and $\pi$-junction phases. When the tunneling strength is increased higher order processes cause the TRB phase to widen and the double minimum phase appears.

\section{Flux threaded $s_\pm$-$s$ Loop}\label{sec:loop}

We now proceed to use the formalism developed for studying an $s_\pm$-$s$ junction to treat the experimentally relevant problem of a flux threaded $s_\pm$-$s$ loop. The system we study consists of bending the $s_\pm$-$s$ junction along the $x$-direction and adding another planar contact to form a loop. Each of the contacts forming the loop is characterized by two tunneling parameters: $w_\alpha^{(1(2))}$, which describe the amplitude of tunneling an electron from the $\alpha$ orbital in the $s_\pm$ superconductor to an adjacent site in the $s$-wave superconductor across contact $1(2)$.

After the addition of the second contact, the Hamiltonian for the $s_\pm$-$s$ junction can still be described by Eq.~(\ref{eqn:juncHamiltonian}) if we modify $T_\alpha$ as:
\begin{equation}
\MatEl{T_\alpha} = -w_\alpha^{(1)} \delta_{m,N_1} \delta_{n,1} -w_\alpha^{(2)} \delta_{m,1} \delta_{n,N_2}
\end{equation}
where the sites in the $s_\pm$ side of the loop are enumerated $1..N_1$ and the sites on the $s$-wave side are enumerated $1..N_2$. The self-consistency equations remain those given in Eq.~(\ref{eqn:sc}).

Next, we proceed by threading the loop with a magnetic flux $\Phi$. As shown in the Appendix, the flux dependence can be transferred to the contact by performing a gauge transformation. With the addition of magnetic flux the tunneling matrices become:
\begin{equation}
(T_\alpha)_{m,n} = -w_\alpha^{(1)} e^{ i\frac{\phi_1}{2}} \delta_{m,N_1} \delta_{n,1} -w_\alpha^{(2)}e^{ i\frac{\phi_2}{2}} \delta_{m,1} \delta_{n,N_2}
\end{equation}
Here, $\frac{\phi_1}{2}$ is the phase an electron acquires through a clockwise hopping across the tunneling contact $1$ and $\frac{\phi_2}{2}$ the phase acquired by an anti-clockwise hopping across the contact $2$. These two phases will fully account for the effect of the magnetic flux as long as $\phi_1$ and $\phi_2$ obey:
\begin{equation}
\phi_1 - \phi_2 = \frac{2 \pi \Phi}{\Phi_0},
\label{eqn:phaseseqn}
\end{equation}
where $\Phi_0 = \frac{hc}{2\mathrm{e}}$ is the superconducting flux quantum.

For long loops, $N_1,N_2 \gg 1$ the two junctions essentially decouple. The OPs behavior near one contact is uninfluenced by the presence of the other contact and the energy cost of the two contacts is a simple sum.
Moreover, by making an additional gauge transformation, the phase acquired by hopping between the two superconductors, $\frac{\phi_{1(2)}}{2}$, can be translated into a phase difference $\phi_{1(2)}$ between the two superconductor OPs. 
Hence, the ground state energy of the loop is given by: 
\begin{equation}
E \left( \Phi \right) = \min_{\phi_1-\phi_2 = \frac{2\pi \Phi}{\Phi_0}} \left( E_1\left(\phi_1\right) + E_2\left(\phi_2\right)\right),
\label{eqn:min}
\end{equation}
where $E_{1(2)} (\phi_{1(2)})$ is the energy of a single $s_\pm$-$s$ junction where the phase difference between the superconductors is $\phi_{1(2)}$ with contact parameters $w_\alpha^{(1(2))}$.

If we denote the phase difference that minimizes the energy of junction 1(2) by $\phi_{1}^{min}$($\phi_{2}^{min}$) then the values of flux that minimize $E(\Phi)$ are given by $\pm \frac{\Phi_0}{2\pi} \left( \phi_{1}^{min}+\phi_{2}^{min} \right)$, $\pm \frac{\Phi_0}{2\pi}\left(\phi_{1}^{min}-\phi_{2}^{min} \right)$. This allows us to deduce the energy of the flux threaded loop as a function of the flux.
\begin{figure}
\centering
\subfigure[\ 0-loop ]{\includegraphics[width=0.48\linewidth]{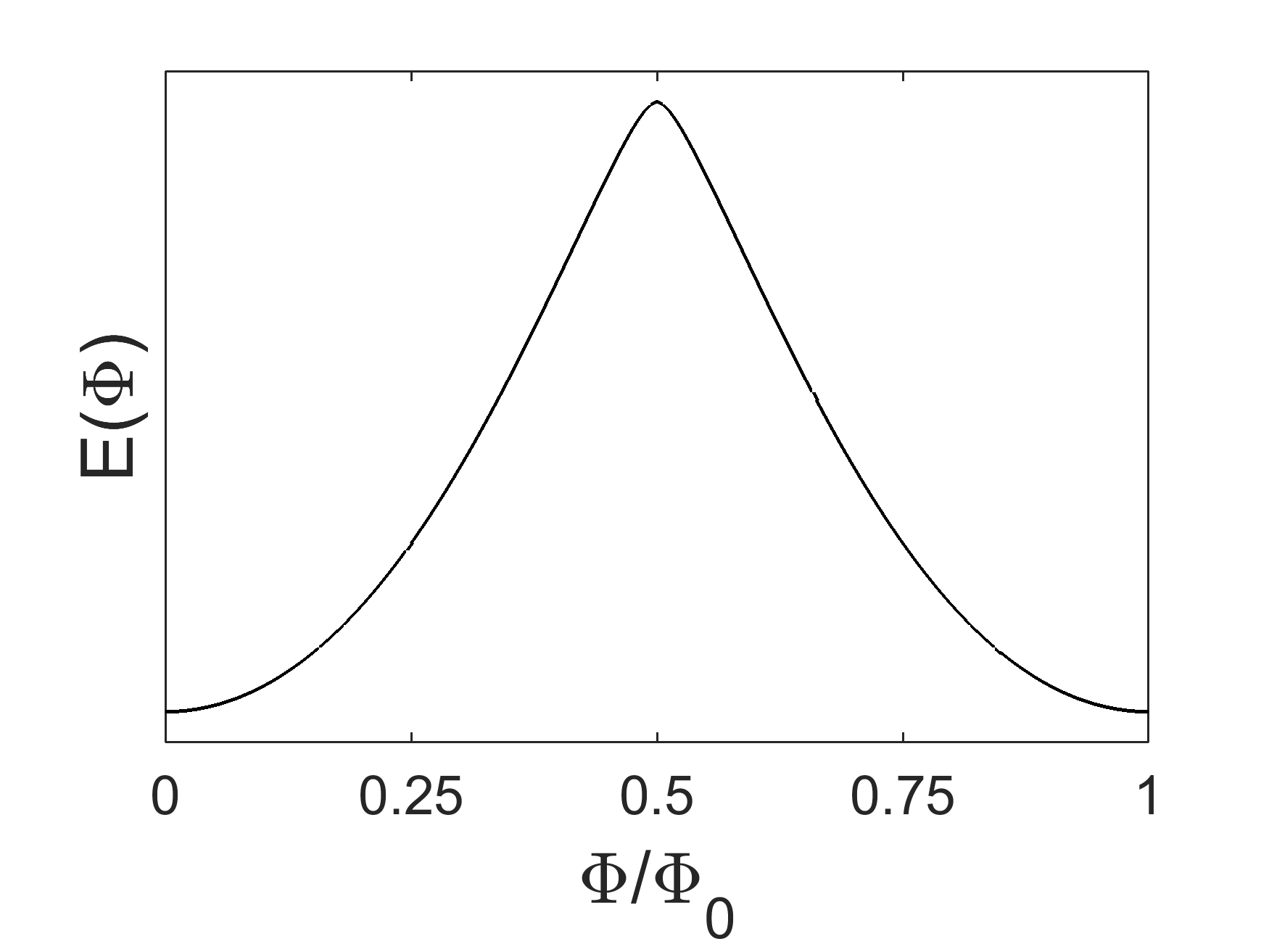}\label{fig:0loop}}
\subfigure[\ $\pi$-loop ]{\includegraphics[width=0.48\linewidth]{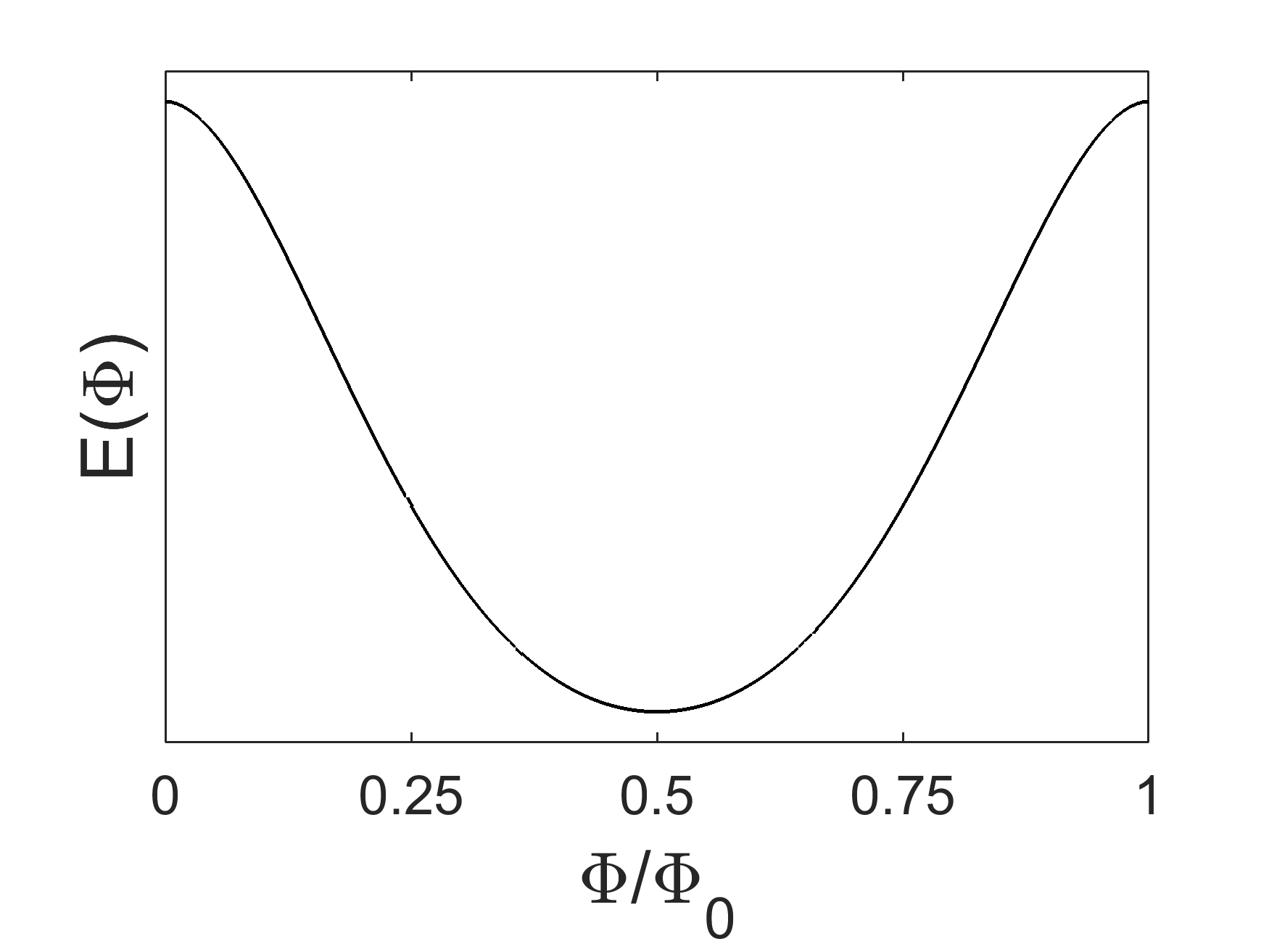}\label{fig:piloop}}
\subfigure[\ TRB ]{\includegraphics[width=0.48\linewidth]{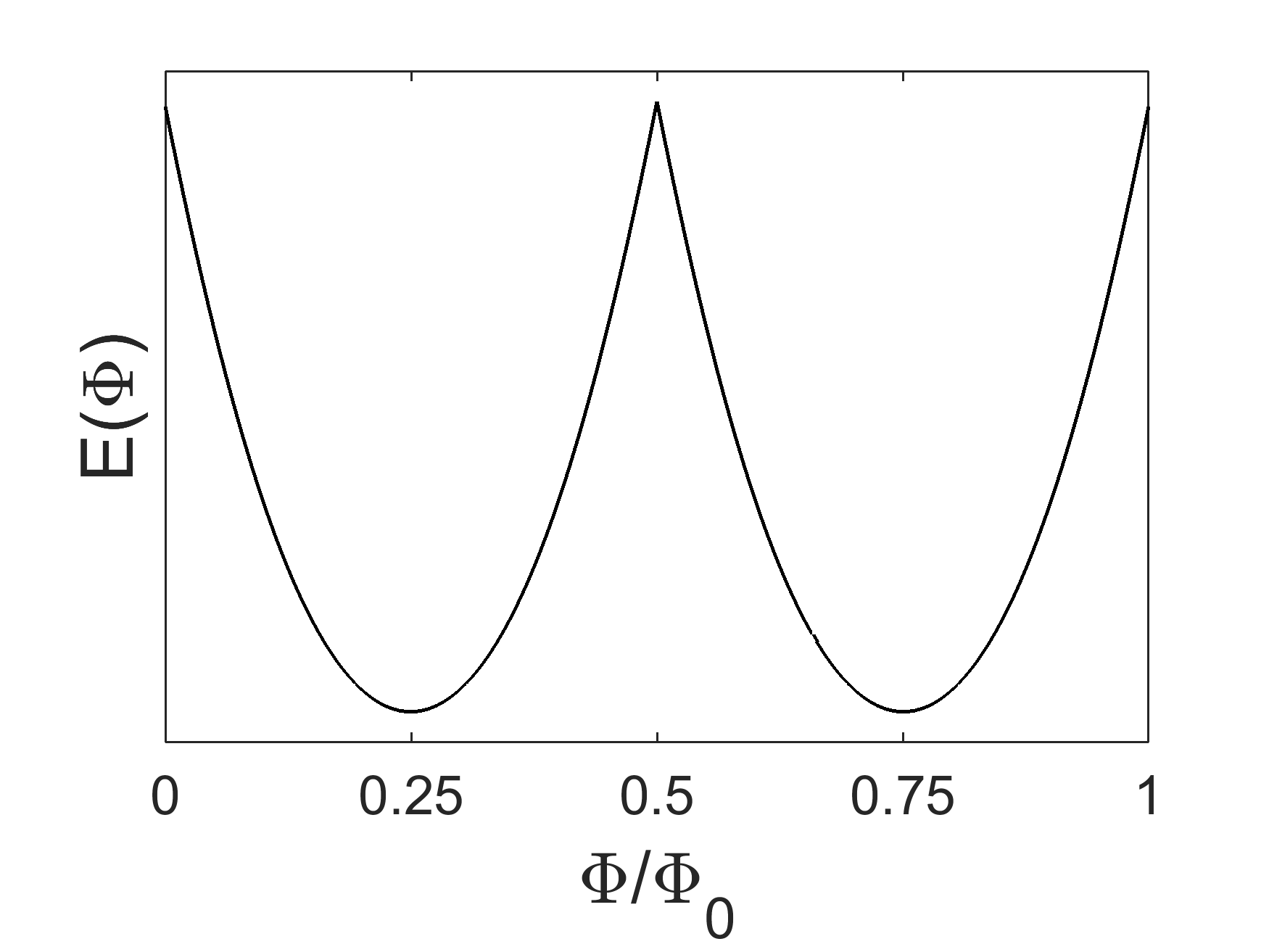}\label{fig:TRB1}}
\subfigure[\ TRB']{\includegraphics[width=0.48\linewidth]{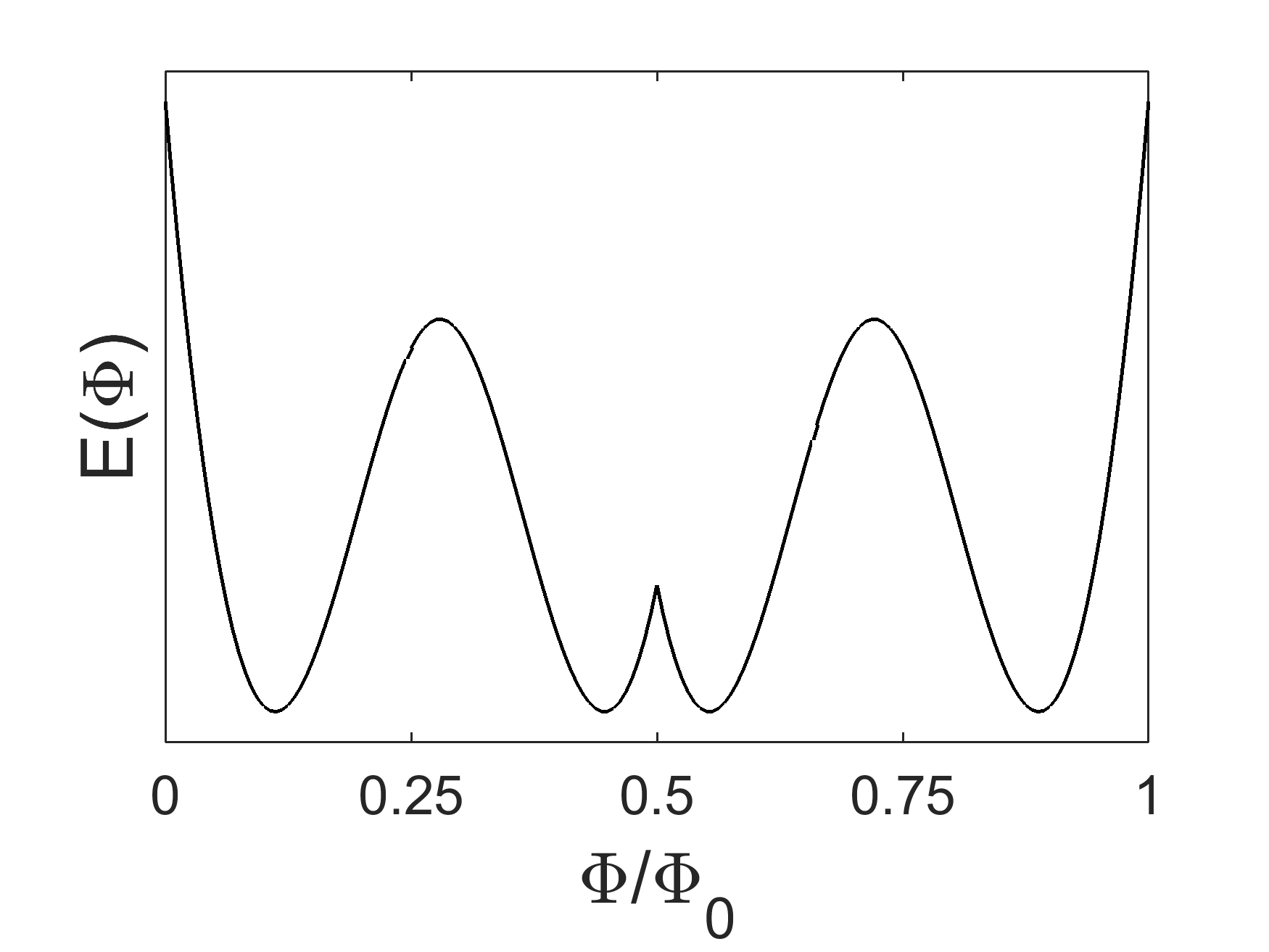}\label{fig:TRB2}}
\subfigure[\ Metastable ]{\includegraphics[width=0.48\linewidth]{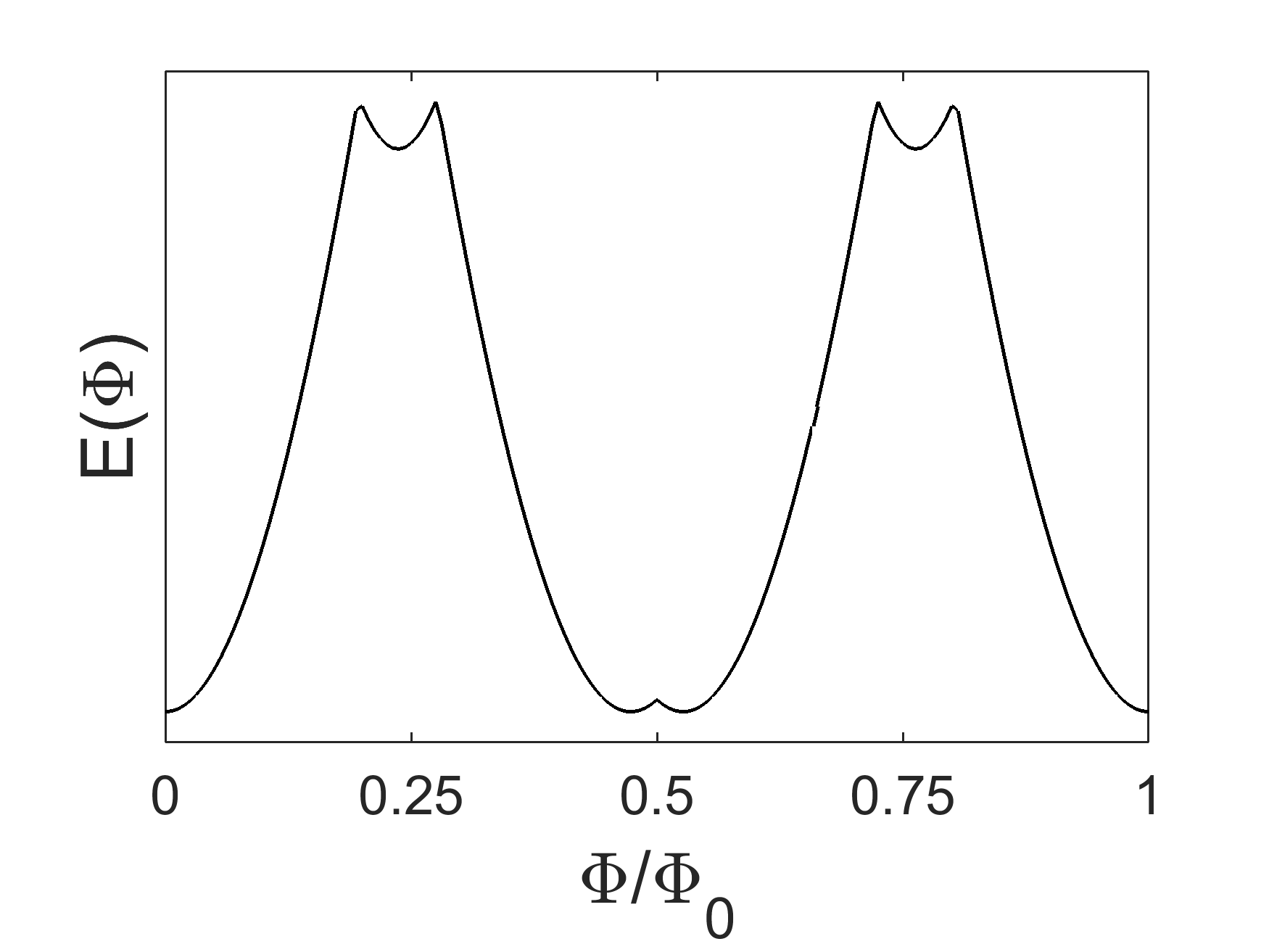}\label{fig:Metastable}}
\subfigure[\ Metastable' ]{\includegraphics[width=0.48\linewidth]{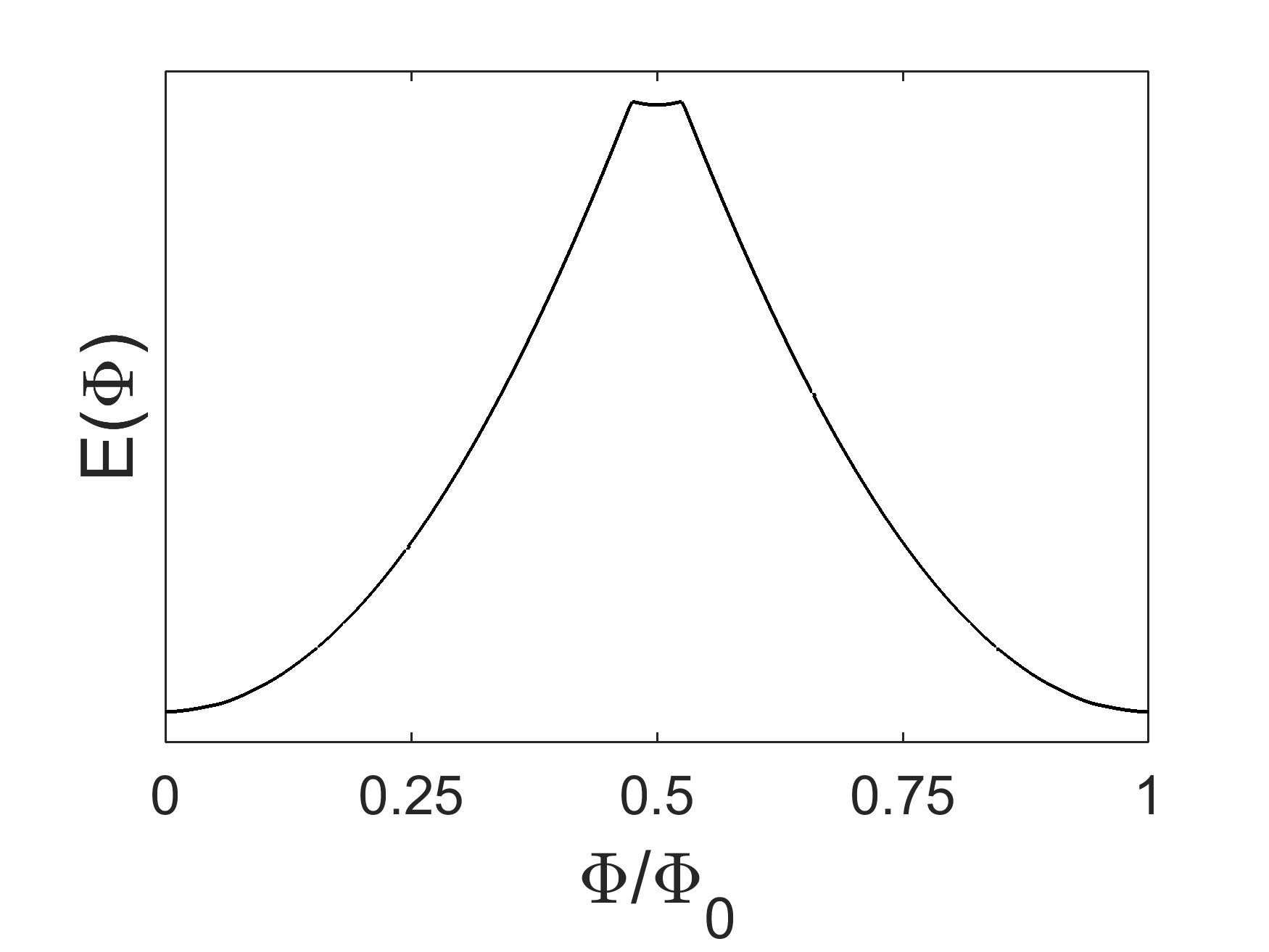}\label{fig:Metastable'}}
\caption{\label{fig:fluxexamples} Energy vs flux examples. In panels (a)-(e), the $s$-wave superconductor parameters are $\Delta_0=0.04$, $\mu_0=-1.3$ and $U=1.446$, while the junction parameters are: (a) $w_x^{(1)}=w_y^{(1)}=0.146$, $w_x^{(2)}=0.31$, and $w_y^{(2)}=0.092$, (b) $w_x^{(1)}=w_y^{(1)}=w_x^{(2)}=0.146$ and $w_y^{(2)}=0.8$, (c) $w_x^{(1)}=w_y^{(1)}=w_x^{(2)}=0.146$ and $w_y^{(2)}=0.364$, (d) $w_x^{(1)}=w_y^{(1)}=0.31$ and $w_x^{(2)}=w_y^{(2)}=0.637$, and (e) $w_x^{(1)}=w_x^{(2)}=0.5$ and $w_y^{(1)}=w_y^{(2)}=0.528$. In panel (f), $\Delta_0=0.02$, $\mu_0=0.3438$, $U=0.9341$, $w_x^{(1)}=w_y^{(1)}=0.6638$, and $w_x^{(2)}=w_y^{(2)}=0.6910$.}
\end{figure}

We find four types of behavior of the energy vs. flux curve, shown in Fig.~\ref{fig:fluxexamples}: 1) Integer-flux-loop with energy minima at integer flux quantum values, 2) $1/2$-flux-loop with energy minima at half integer flux quantum values, 3) Time reversal broken loop (TRB) with energy minima at fractional values of flux quantum and 4) Meta-stable, with two types of energy minima. While the first three cases have been found by several authors\citep{trijunctionpointcontact,secondharmonicballistic,
secondharmonicballisticprevious,GolubovandMazin,
ExperimentProposal_BarierThickness,cornerjunctionmodification,Microscopic2,Lagrangian,TRBGL1,TRB2}, the possibility of an energy/flux relation with minima of different depth has not been seen in a microscopic theory before.

The case where the energy of the loop is minimized for integer values of flux quantum occurs whenever the energy of both of the contacts is minimized at a phase difference of $0$ or $\pi$.

The energy of the loop is minimized for half-integer flux quantum values when the energy/phase relation one of the contacts has a minimum at $0$ and the other has a minimum at $\pi$. Our analysis shows that it is possible to find this behavior without changing the bulk parameters of the superconductors if the tunneling parameters across each of the two contacts are different.

If the energy of one of the contacts is minimized for a phase difference $0<\phi<\pi$, then the energy of the loop will be minimized for values of magnetic flux which are neither integer flux quantum nor half integer flux quantum. This causes supercurrent to flow in the loop and therefore the phase is named `time reversal breaking'. The energy-flux relation shown in Fig.~\ref{fig:TRB1} results from having one of the contacts in the $\phi$-junction phase, while the other is in the $0(\pi)$-junction phase. On the other hand, if the energy of both contacts is minimized for a phase difference $0<\phi<\pi$ we obtain four degenerate minima: $\pm \frac{\Phi_0}{2\pi} \left( \phi_{1}^{min}+\phi_{2}^{min} \right)$, $\pm \frac{\Phi_0}{2\pi}\left(\phi_{1}^{min}-\phi_{2}^{min} \right)$. An example of this type of energy vs flux relation is shown in Fig.~\ref{fig:TRB2}. For a significant portion of the $\phi$-junction phase, the value of the minimum is close to $\pi/2$.
If this is the case for the two junctions forming the loop, we will find two energy minima close to integer flux quantum and two energy minima close to half integer flux quantum.

An energy/flux relation with minima of different depth such that one is a global minimum and the other is a (meta-stable) local minimum can occur when the loop is formed by two contacts that have a double minimum. As can be seen in Figure \ref{fig:Metastable}, the energy/flux relation exhibits the four degenerate minima of Fig.~\ref{fig:TRB2} and two additional local minima. If one of the contacts is in a double minimum phase while the other is not, the resulting energy-flux behavior will be that of Fig.~\ref{fig:TRB1} or of Fig.~\ref{fig:TRB2}. In other words, in order to detect signatures of the double minimum regime in the energy-flux relation, both contacts must be in this regime. Therefore to obtain a loop with minima of different depth, it is necessary to have a) a large tunneling amplitude across the two contacts and b) a large pairing amplitude in the $s$-wave superconductor.

\begin{figure}
\centering
\subfigure[\ Metastable ]{\includegraphics[width=0.48\linewidth]{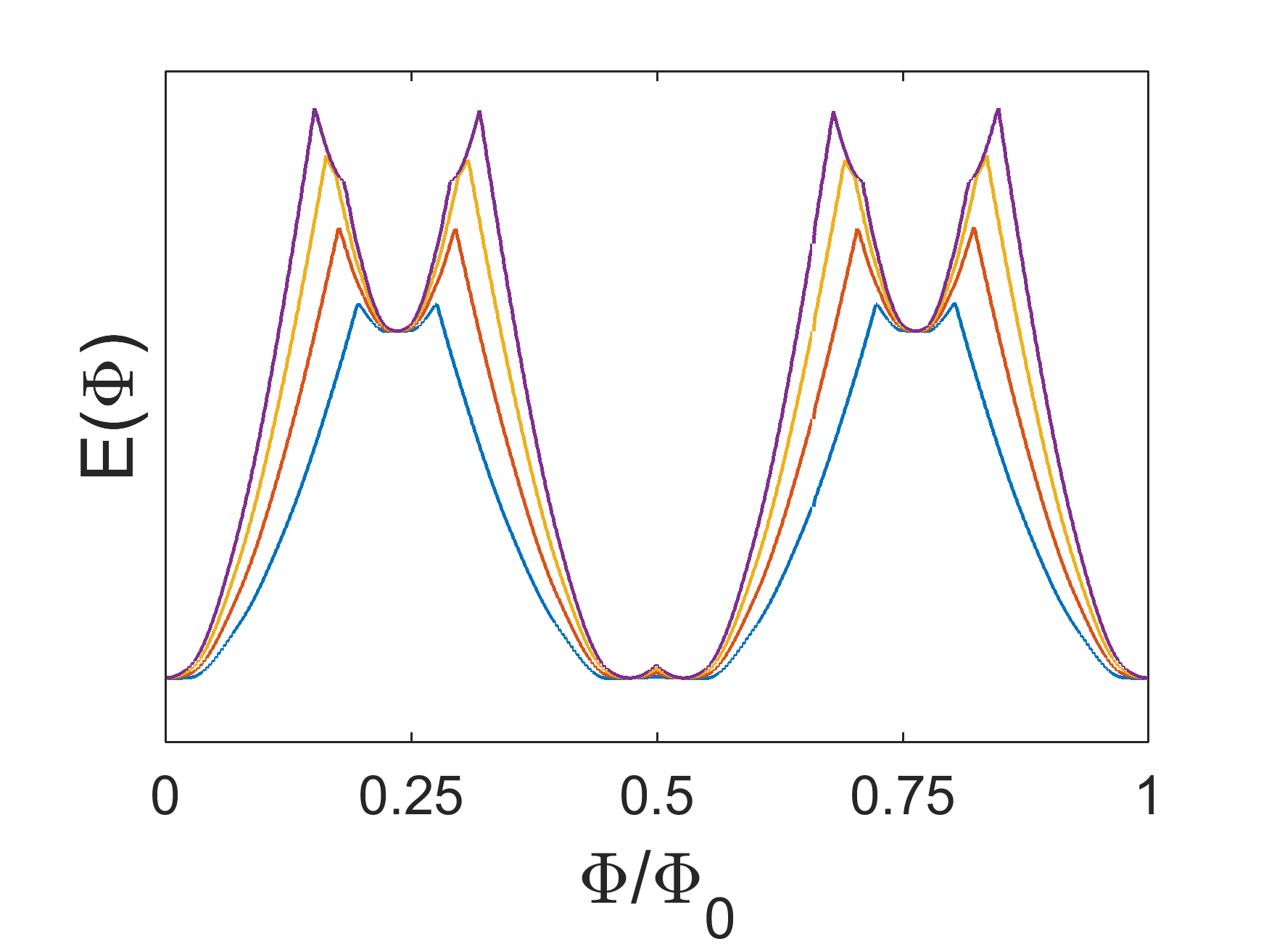}}
\subfigure[\ Metastable' ]{\includegraphics[width=0.48\linewidth]{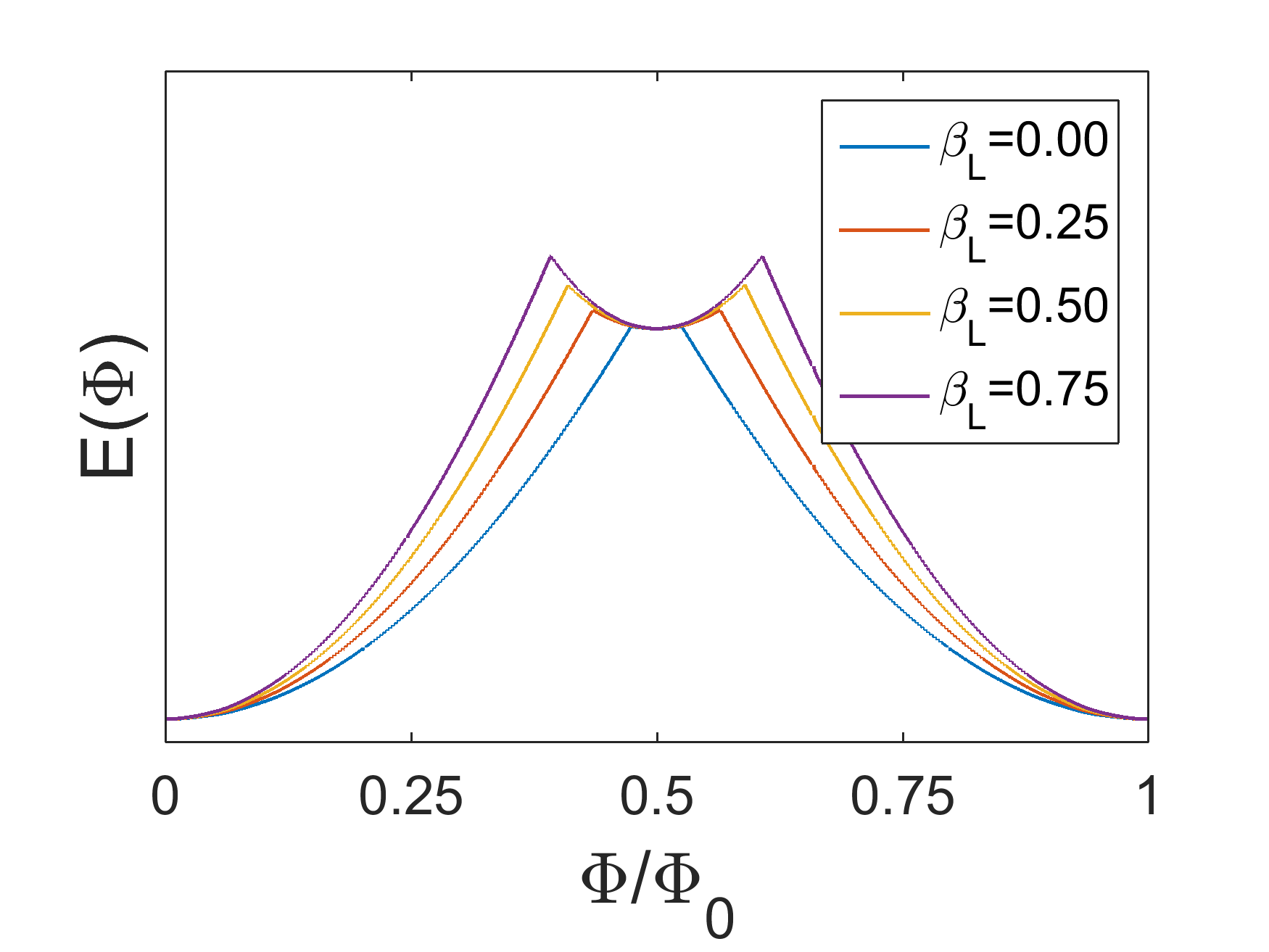}}
\caption{\label{fig:Leffects} Effects of inductance in the Metastable and Metastable' energy vs flux relations. }
\end{figure}

The local minima in the meta-stable relations found in this work are shallow (see Fig.~\ref{fig:Metastable} and Fig.~\ref{fig:Metastable'}) and hence they would be very hard to detect experimentally. However, we find that these minima are considerably deeper for loops with large inductance. The ground state energy of a loop threaded by a total flux $\Phi$ and with inductance $L$ is given by\citep{SQUID}:
\begin{equation}
\begin{split}
E \left( \Phi \right) = \min_{\phi_1-\phi_2 = \frac{2\pi \Phi}{\Phi_0}} \left( \sum_{i=1,2} \left( E_i\left(\phi_i\right) + \frac{L}{4}I_i^2(\phi_i) \right) \right)
\end{split}
\label{eqn:min2}
\end{equation}
where for simplicity we have considered an equal inductance in both arms of the loop. In Fig. \ref{fig:Leffects}, we show how the energy vs flux relations of Fig.~\ref{fig:Metastable} and Fig.~\ref{fig:Metastable'} are modified for different values of the screening parameter $\beta_L = \frac{L(I_{0,1} + I_{0,2})}{\Phi_0}$, where $I_{0,1}$ and $I_{0,2}$ are the critical currents of the two junctions. As can be seen in Fig. \ref{fig:Leffects}, increasing the inductance of the loop has the effect of deepening the local energy minima.

\section{Conclusions}
We studied the Josephson tunneling between an $s_\pm$ superconductor and a single band $s$-wave superconductor within a microscopic formalism in which the order parameters of both superconductor are determined using self-consistent Bogoliubov-deGennes equations. We find four possible junction behaviors, characterized by their energy/phase difference relation. The possible states are: (i) 0-junction where the energy minimum is at zero phase difference, (ii) $\pi$-junction where the energy minimum occurs at $\pi$, (iii) $\phi$-junction where $0<\phi<\pi$, and (iv) a double minimum junctions where there are two minima, one of them global and the other local.

We find that allowing the order parameter to change its amplitude and phase self-consistently close to the junction has some important effects on the resultant phase diagram. Most notable - it is essential for the appearance of a new state, namely the double minimum junction.

We also use our results to study the energy of a flux threaded $s_\pm$-$s$ loop. We find that the loop can have different types of unconventional energy/flux behavior such as, $1/2$-flux-loop, TRB and metastable.

\begin{acknowledgments}
R.R. acknowledges financial support from the FQRNT through Merit Scholarship Program for Foreign students and from the Secretary of Public Education and the Government of Mexico. T.P. acknowledges financial support from the NSERC and the FQRNT. E. B. was supported by the ISF under grant 1291/12, by the US-Israel BSF, and by a Minerva ARCHES prize. Numerical calculations for this work were performed using Calcul Qu\'{e}bec computing resources.
\end{acknowledgments}

\bibliographystyle{apsrev}
\bibliography{summary4}
\appendix
\section{Magnetic flux}
\label{sec:Magneticflux}
In this Appendix we explain how the magnetic flux is added to our microscopic model.
For non-zero magnetic flux, we select a gauge in which the magnetic potential is parallel to the angular direction, i.e. $\textbf{A} = A\hat{x}$. The magnetic flux enclosed by the loop $\Phi$, must be equal to $\oint \textbf{A} \cdot d \textbf{l}$. This yields $\textbf{A} = \frac{\Phi}{N_1+N_2} \hat{x}$, since the diameter of the circle is equal to the number of sites in the $x$-direction times the lattice constant $\left(N_1+N_2\right)a$ and we have set $a=1$. The hopping terms in the Hamiltonian are modified with the introduction of the magnetic potential according to Peierls substitution.
\begin{equation}
t \rightarrow t \exp\left(- \frac{i\mathrm{e}}{\hbar c} \int_{r'}^{r} \textbf{A} \cdot d \bf{l}\right)
\end{equation}

For nearest neighbor hopping around the loop this leads to $t \rightarrow t e^{-i\phi}$ with $\phi$ given by:
\begin{equation}
\phi = \frac{\pi}{N_1+N_2} \left(\frac{\Phi}{\Phi_0}\right)
\end{equation}
where $\Phi_0 = \frac{hc}{2\mathrm{e}}$ is the superconducting flux quantum. The system can still be described by equation (\ref{eqn:juncHamiltonian}) with the appropriate modification of the matrices $K_{s_\pm}$, $K_s$, $T_x$ and $T_y$ which become dependent on the phase $\phi$.

In order to simplify the Hamiltonian we use the transformation $\destruction{c}{k_y,\sigma}(n) \rightarrow e^{i \frac{\phi_1}{2} - i n \phi} \destruction{c}{k_y,\sigma}(n)$, $\destruction{d}{\alpha,k_y,\sigma}(n) \rightarrow e^{-i n\phi}\destruction{d}{\alpha,k_y,\sigma}(n)$, which accordingly modifies the self consistent order parameters defined by Eq.~(\ref{eqn:sc}) as $\Delta_0 \left(n\right) \rightarrow e^{i\phi_1 - 2 i n \phi} \Delta_0 \left(n\right)$ , $\Delta_\alpha \left(n,n+1\right) \rightarrow e^{- i(2n+1)\phi} \Delta_\alpha \left(n,n+1\right)$.

It is easy to see that under this transformation the flux dependence is transferred to the contacts. This can be seen in the transformed Hamiltonian with the definitions in Eq.~(\ref{eqn:juncHamiltonian}) as $K_{s_\pm}$ and $K_s$ loose their dependence and the pairing matrices $\Delta_{s_\pm}$ and $\Delta_s$ are invariant while the $T_{\alpha}$ matrices are now given by:
\begin{equation}
(T_\alpha)_{m,n} = -w_\alpha^{(1)} e^{ i\frac{\phi_1}{2}} \delta_{m,N_1} \delta_{n,1} -w_\alpha^{(2)}e^{ i\frac{\phi_2}{2}} \delta_{m,1} \delta_{n,N_2}
\end{equation}
where $\phi_1$ and $\phi_2$ are solutions of
\begin{equation}
\phi_1 - \phi_2 = \frac{2 \pi \Phi}{\Phi_0}
\end{equation}

\section{Numerical Methods}
\label{sec:num}

\subsection{Finding the relation between the superconducting coupling constants and the bulk pairing amplitudes}

In this work, we set the bulk pairing amplitudes $\Delta_x(n,n+1) = \Delta_y(n,n+1)=0.081$. In a translationally invariant system, this corresponds to the Hamiltonian given by equations (\ref{eqn:spmMF}) and (\ref{eqn:spmParams}) with $\Delta_{x} = \Delta_{y} = 0.162$. The value of $J_2$ is then set such that the solution of the self-consistency equation (\ref{eqn:SCspm}) for $\qv = 0$ is $\Delta_{x} = \Delta_{y} = 0.162$. The self-consistent equation (\ref{eqn:SCspm}) can be solved numerically, iterating from an initial guess. In a 100$\times$100 lattice, the coupling $J_2 = 0.624$ leads to the desired value of the bulk $s_\pm$ pairing amplitude. The iteration loop was stopped when the difference between the input and the calculated order parameters was less than $1\times10^{-5}$. 
\begin{figure}
\subfigure[\ ]{\includegraphics[width=0.48\linewidth]{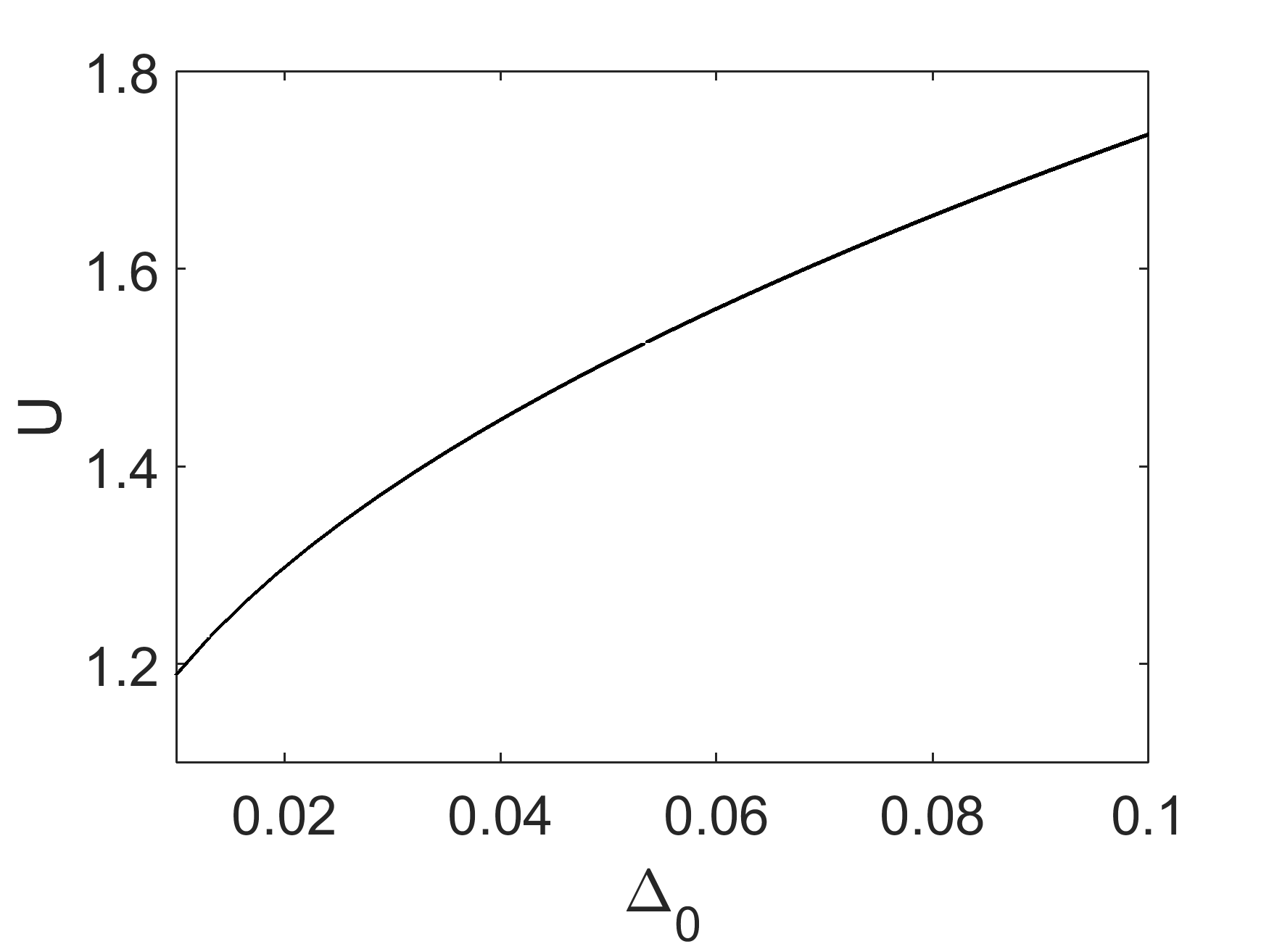}}
\subfigure[\ ]{\includegraphics[width=0.48\linewidth]{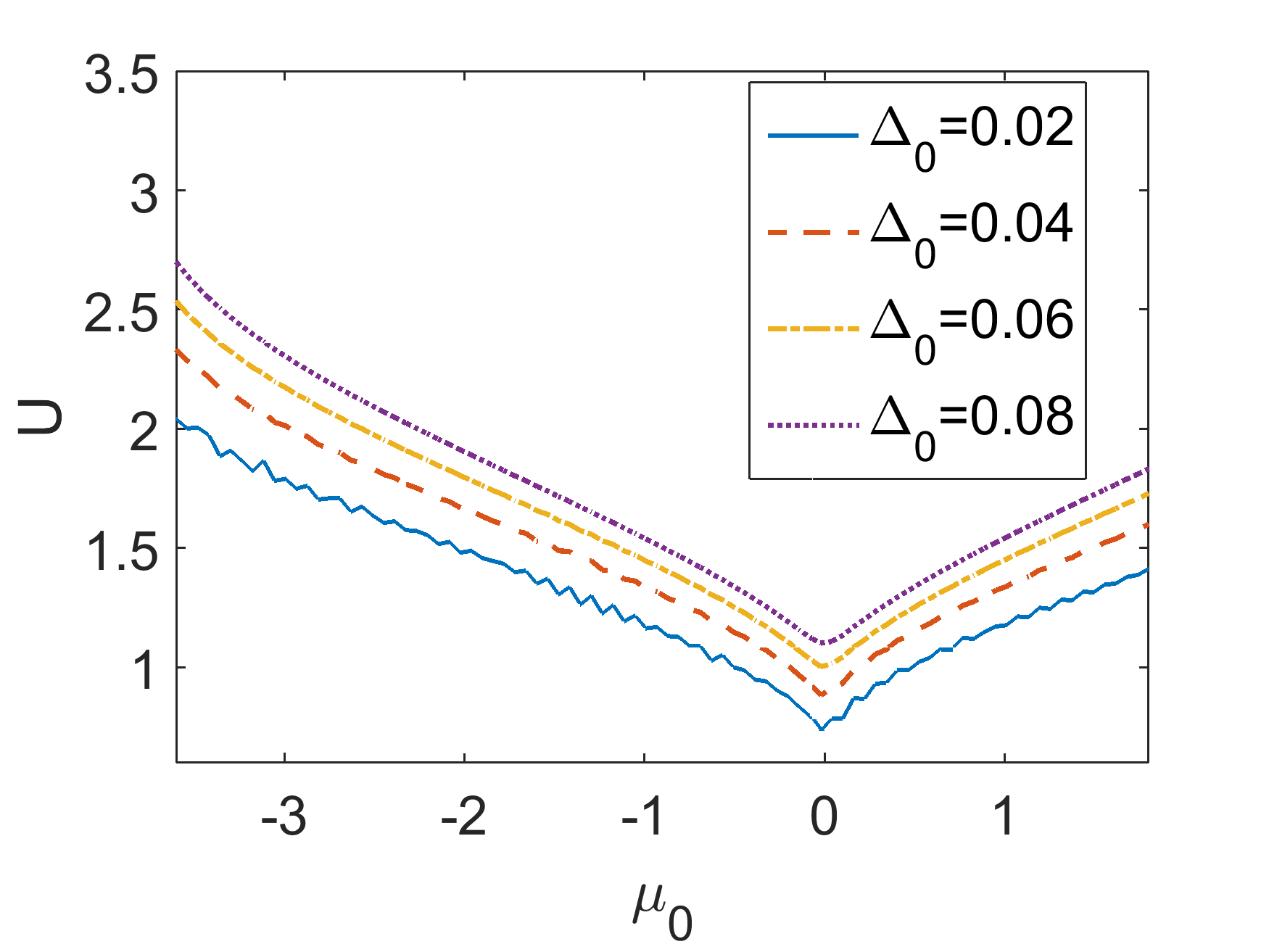}}
\caption{\label{fig:Uvals} Values of the superconducting coupling $U$ used in (a) Fig. \ref{fig:sc13} and (b) Fig. \ref{fig:tvsmu}.}
\end{figure}

The bulk pairing amplitude $\Delta_0$ corresponding to given values of $\mu_0$ and $U$ can be calculated by solving the self-consistency equation (\ref{eqn:SCs}). This was done iterating equation (\ref{eqn:SCs}) in a 100$\times$100 lattice with periodic boundary conditions until the input and the calculated order parameters was less than $1\times10^{-5}$. To obtain the value of $U$ corresponding to given values of $\mu_0$ and $\Delta_0$, we used a numerical solver to solve the equation $\Delta_0(\mu_0,U)-\Delta_0=0$ with the function $\Delta_0(\mu_0,U)$ defined as the (numerical)solution of the self-consistency equation. The values of $U$ obtained for Figures \ref{fig:sc13} and \ref{fig:tvsmu} are shown in Fig. \ref{fig:Uvals}, some finite size effects can be appreciated in the solution for $\Delta_0 =0.02$. \\

\subsection{Self-consistent solution of the BdG equations - calculating the order parameter magnitude close to the junction}

The self-consistency equations \ref{eqn:sc} were solved in the vicinity of the interface, using the bulk values of the order parameters, i.e $\Delta_x (n,n+1) = \Delta_y (n,n+1) = 0.081 $ and $\Delta_0(n) = |\Delta_0| e^{i \phi}$, as a starting point of the iteration loop. The iteration loop is stopped when the difference in the order parameters obtained in two consecutive iterations is less than $1\times10^{-6}$. We consider periodic boundary conditions on the $y$-direction and a $20\times20$ lattice for each superconductor.

When changing the momentum resolution in the direction along the junction cross section ($k_y$) we find some sensitivity of our results to the resolution. However, the main findings are not altered. In the phase diagram the nature of the different phases is not sensitive to the $k_y$ resolution but the phase boundaries may shift slightly.

\subsection{Energy/flux curves}

The energy/flux relation for the array in Sec.~\ref{sec:loop} can be easily found from the energy/phase profile of the two interfaces following Eqn.~\ref{eqn:min}. The energy/~phase relation is found by solving the order parameter self-consistently and the energy using exact diagonalization for 41 evenly spaced values of the phase difference between $0$ and $2\pi$. Afterwards, we can define the energy/phase relations $E_{1(2)}(\phi)$ for any value of $\phi$ using cubic Hermite spline interpolation. Finally, the energy vs flux curve is given by:

\begin{equation}
E \left( \Phi \right) = \min_{0 \leq \phi \leq 2\pi} E_1 \left( \phi +\frac{2\pi \Phi}{\Phi_0} \right) + E_2\left(\phi \right),
\end{equation}

Since the value of $\phi$ is bounded and the minima of $ E_1 \left( \phi +\frac{2\pi \Phi}{\Phi_0} \right) + E_2\left(\phi \right)$ are very sharp for the metastable cases, the most practical way to do the minimization is by brute force, i.e. by directly evaluating the function for a grid of points in the $[0,2\pi]$ interval. The curves is Fig. ~\ref{fig:fluxexamples} were obtained using a grid of $1000$ points.

\end{document}